\documentclass[twocolumn,superscriptaddress,aps,amsmath,amssymb,prb]{revtex4}
\usepackage{bm}
\usepackage{setspace}
\usepackage{graphicx}
\usepackage{graphics}

\newcommand{\Sec}[1]{Sec.\,\ref{#1}}
\newcommand{\App}[1]{Appendix\,\ref{#1}}

\newcommand{\nl}{\nonumber \\}
\newcommand{\ie}{{\rm i.e.\ }}

\newcommand{\etal}{{\it et al.\ }}
\newcommand{\be}{\begin{equation}}
\newcommand{\ee}{\end{equation}}
\newcommand{\bea}{\begin{eqnarray}}
\newcommand{\eea}{\end{eqnarray}}
\newcommand{\bsube}{\begin{subequations}}
\newcommand{\esube}{\end{subequations}}
\newcommand{\Fig}[1]{Fig.\,\ref{#1}}
\newcommand{\Eq}[1]{Eq.\,(\ref{#1})}
\newcommand{\Eqs}[1]{Eqs.\,(\ref{#1})}

\newcommand{\ep}{\epsilon}

\newcommand{\w}{\omega}
\newcommand{\eps}{\epsilon}

\newcommand{\dd}{_{\rm D}}
\newcommand{\tlam}{\tilde{\bm\Lambda}}

\newcommand{\trt}{{\rm tr_{_T}}}

\begin{document}

\title{Time--dependent density functional theory for quantum transport}

\author{Xiao Zheng}
\affiliation{Department of Chemistry, The Hong Kong University
      of Science and Technology, Hong Kong}
\affiliation{Department of Chemistry, The University of Hong Kong,
Hong Kong}

\author{GuanHua Chen}
\email{ghc@everest.hku.hk} \affiliation{Department of Chemistry,
  The University of Hong Kong, Hong Kong}
\affiliation{Department of Physics, The University of Hong Kong,
 Hong Kong}
\affiliation{Centre for Theoretical and Computational Physics, The University of Hong Kong, Hong Kong}

\author{Yan Mo}
\affiliation{Department of Chemistry, The Hong Kong University
      of Science and Technology, Hong Kong}
\affiliation{Department of Chemistry, The University of Hong Kong,
 Hong Kong}
\affiliation{Department of Applied Chemistry, National Chiao Tung
      University, Hsinchu 30050, Taiwan}

\author{SiuKong Koo}
\author{Heng Tian}
\author{ChiYung Yam}
\affiliation{Department of Chemistry, The University of Hong Kong,
 Hong Kong}

\author{YiJing Yan}
\email{yyan@ust.hk}\affiliation{Department of Chemistry, The Hong Kong University
      of Science and Technology, Hong Kong}

\date{\today}

\begin{abstract}

Based on our earlier works [Phys.\ Rev.\ B \textbf{75}, 195127
(2007) \& J.\ Chem.\ Phys.\ \textbf{128}, 234703 (2008)], we propose
a formally exact and numerically convenient approach to
simulate time--dependent quantum transport from first--principles.
The proposed approach combines time--dependent
density functional theory with quantum dissipation theory, and
results in a useful tool for studying transient dynamics of
electronic systems. Within the proposed exact theoretical framework,
we construct a number of practical schemes for simulating realistic
systems such as nanoscopic electronic devices. Computational cost of
each scheme is analyzed, with the expected level of accuracy
discussed. As a demonstration, a simulation based on the adiabatic
wide--band limit approximation scheme is carried out to characterize
the transient current response of a carbon nanotube based electronic
device under time--dependent external voltages.

\end{abstract}

\maketitle

\section{Introduction}
\label{tddft-open}

The combined density functional theory (DFT) and nonequilibrium
Green's function (NEGF) approach has been widely employed to
simulate steady state quantum electron transport through molecular
junctions and other nanoscopic structures. In practice, DFT
introduces an effective single--electron reference system, on which
NEGF analysis for steady current and electron occupation can be
worked out. However, in principle it remains obscure how the
conventional DFT for ground/equilibrium state of isolated systems
can provide a rigorous framework for quantum transport, as electron
transport is intrinsically a dynamic process.

Time--dependent density functional theory (TDDFT)\cite{gross} has
been developed to study quantum transport
phenomena.\cite{stef,kurt,zheng1,zheng2,zheng3,yan1,Jin08234703,burke,chen1}
As a formally rigorous and numerically tractable approach, TDDFT
promises real--time simulations on ultrafast electron transport
through realistic electronic devices or structures.
In early attempts, finite source--device--drain systems were treated by the
conventional TDDFT for isolated systems.\cite{aiichiro,voor,ventra}
Since the source and drain were treated as finite, the results were
not directly transferable to realistic devices coupled to bulk electrodes.
By employing the NEGF approach and a partition--free scheme,
Stefanucci and Almbladh have derived the exact equations of motion
for the two--time Green's functions within TDDFT
framework.\cite{stef} They and their coworkers have further proposed
a practical scheme in which the electronic wavefunction propagates
in time domain subject to open boundaries.\cite{kurt} The resulting
numerical method has been tested on model systems. However, its
applicability to realistic electronic devices remains unexploited.

Based on a reduced single--electron density matrix (RSDM) based
formulation,\cite{ldm1, ldm2} we have proposed a rigorous TDDFT
approach for open electronic systems.\cite{zheng1,chen1} We have
also developed an accurate numerical scheme,\cite{zheng2} based on a
closed equation of motion (EOM) for the Kohn--Sham (KS) RSDM. Our
RSDM based TDDFT--EOM has been combined with the Keldysh's
non--equilibrium Green's Function (NEGF) formalism, the resulting
TDDFT--NEGF--EOM approach has been applied successfully to exploit
transient electronic dynamics in realistic molecular electronic
devices.\cite{zheng2} One of recent applications of the
TDDFT--NEGF--EOM method was a simulation on the ultrafast transient
current through a carbon nanotube based electronic device. It was
found that the dynamic electronic response of the device can be
mapped onto an equivalent classical electric
circuit,\cite{zheng3,Mo09355301} which would be useful for future
design of functional devices.

Our previous simulations on realistic nanoelectronic
devices\cite{zheng2, zheng3} have demonstrated the numerical
feasibility of our TDDFT--NEGF--EOM approach for open systems, while its
accuracy is largely determined by the quality of approximated
exchange--correlation (XC) functional which accounts for the
many--particle effects, and that of dissipation functional which
characterizes the dissipative interactions between the electronic
device and electrodes. In our previous simulations we employed the
adiabatic local density approximation\cite{Koh651133} (ALDA) for the
XC functional and the adiabatic wide--band limit\cite{zheng2} (AWBL)
approximation for the dissipation functional. Despite the success,
the ALDA approximation is expected to become inadequate for
addressing transient dynamics of open electronic systems in
circumstances where electron correlations dominate.
Moreover, the frequency dispersion part of
the XC potential is completely missing from an adiabatic XC
functional such as ALDA,\cite{sai} which may lead to loss of crucial
transient features in the electronic dynamics.
The AWBL approximation for dissipation functional may lead to
nontrivial errors for electrodes of finite band widths and strongly
inhomogeneous energy bands, or when non--Markovian memory effects
play a significant role (such as in multi--channel conductors).

 Several groups have shown that within TDDFT framework
the steady state current can be formally expressed by
Landauer--B\"{u}ttiker formula, if the steady state can be
reached.\cite{stef,zheng2,Jin08234703} This is important as the
Landauer--B\"{u}ttiker formula offers a convenient way to evaluate
steady state current. However, in principle, it is the XC potential
of TDDFT which explicitly includes frequency--dependent component
that should be used, instead of that of ground state DFT. In
practice, the use of an adiabatic XC functional would lose all the
memory of transient dynamics, and results in the same steady state
current predicted by ground state DFT.
Therefore, it is desirable to go beyond both the ALDA and AWBL
approximations, and to develop more sophisticated XC and dissipation
functionals for better characterization of transient electronic
dynamics in quantum transport systems and structures.

Time--dependent quantum transport has been addressed from the
perspective of open dissipative systems. Rossi, Di~Carlo and Lugli
have developed a single--electron density matrix based Bloch
equation to simulate quantum--transport through mesoscopic
devices.\cite{carlo}
Burke, Car and Gebauer have proposed a single--electron density
matrix based master equation for a current--carrying electronic
system with dissipation to a phonon bath.\cite{burke}
Weiss~\emph{et al.} have proposed an iterative path--integral
approach,\cite{Wei08195316} which is numerically exact but
computationally expensive. My\"{o}h\"{a}nen~\emph{et
al.}\cite{Myo0867001} have extended the Kadanoff--Baym approach to
open interacting systems,\cite{Kad62,Dah08153004} where the roles of
initial correlations and memory effects were highlighted. The simulation of
transient current has also been attempted by many--body quantum
master equation
approaches.\cite{Har06235309,Har05373,Li08075321,Wel0957008,Jin07134113}
Yan and coworkers have started from a master equation based
second--order quantum dissipation theory (QDT), and derived an EOM
for the RSDM within TDDFT framework,\cite{yan1} where the bulk
electrodes are treated as electron reservoirs.
This was followed by the recent establishment of a formally exact
QDT for the dynamics of an arbitrary non--Markovian dissipative
systems interacting with bath surroundings.\cite{Jin08234703} The
exact QDT is formulated in terms of hierarchical equations of motion
(HEOM), and applied to solve time--dependent quantum transport
problems.\cite{Zhe08184112,Zhe08093016,Zhe09124508,Zhe09164708} The
HEOM--QDT is intrinsically a nonperturbative method, and is
constructed to resolve the combined effects of the many--particle
interaction, dissipative coupling strength, and memory time. The
HEOM--QDT is by far the most tractable exact approach
to time--dependent transient current through interacting
electronic systems under arbitrary time--dependent voltage.

In this work, we aim at unifying our TDDFT--NEGF--EOM approach for
open systems with the HEOM--QDT, to
establish a TDDFT--NEGF--HEOM method which is formally exact and
numerically efficient. With the TDDFT--NEGF--HEOM approach, the
dissipation functional readily goes beyond the AWBL approximation,
resulting in systematic improvement on the simulated transient
electronic dynamics. Moreover, the TDDFT--NEGF--HEOM approach is also
applicable to devices operating at finite temperatures.

This paper is organized as follows. In \Sec{eom-sigma-jt} we briefly
review the TDDFT--NEGF--EOM formalism for open electronic systems, and
introduce the TDDFT--HEOM formalism. The two formalisms are then combined
and resulted in a unified TDDFT--NEGF--HEOM formalism. In \Sec{q-exact}, we exploit
formally exact computational schemes to calculate the dissipation
functional with the TDDFT--NEGF--HEOM formalism. In \Sec{q-approx} we
propose three approximate schemes to reduce further the
computational cost. In \Sec{results} a numerical example is
presented, with analysis on the computational efficiency. We
summarize our results in \Sec{summary}.

\section{Density matrix based time--dependent density
functional theory for open systems}
\label{eom-sigma-jt}

\subsection{RSDM based TDDFT--EOM approach for isolated systems}
\label{tddft-iso}

We start with {\it ab initio} many--particle Hamiltonian of an
\emph{isolated} system containing $N$ electrons:
\be
 H(t) = \sum_{i=1}^N \big[-\frac{1}{2}\nabla^2_i + v({\bm r}_i, t)\big]
  + \sum_{i<j}^N \frac{1}{{\bm r}_{ij}},
\ee
where $i$ and $j$ run over all the $N$ electrons, and $v({\bm r}_i,
t)$ is the external potential for the $i\,$th electron due to nuclei
and other electrostatic sources. Atomic units are adopted here and
throughout the rest of manuscript.

The Runge--Gross theorem\cite{gross} establishes the one--to--one
correspondence between the external potential $v(\bm r, t)$ and the
time--dependent electron density function $\rho(\bm r, t)$ of an
isolated system. A time--dependent Kohn--Sham (TDKS) scheme has been
proposed to solve $\rho(\bm r, t)$, and hence all physical
properties of the isolated system. The TDKS scheme follows the time evolution of an
isolated reference system consisting of $N$ noninteracting
electrons. The Hamiltonian of the reference system is as follows:
\be
   \hat{H}(t) = \sum_{i=1}^N \hat{h}(\bm r_i, t) = \sum_{i=1}^N
   \big[-\frac{1}{2} \nabla^2_i + v_{\rm eff}(\bm r_i, t) \big],
   \label{h_tdks}
\ee
where $v_{\rm eff}(\bm r_i, t)$ is the effective single--electron
potential. A closed EOM has been derived for the KS RSDM $\bm
\sigma$ of the isolated system:\cite{ldm1,ldm2}
%
\be
   i \dot{\bm \sigma}(t) = \left[ \bm h(t), \bm \sigma(t) \right].
   \label{eom-full}
\ee
Here, $\bm h(t)$ is the KS Fock matrix. Its matrix element is
evaluated via $\bm h_{\mu\nu}(t) = \int \chi_\mu(\bm r)
[-\frac{1}{2} \nabla^2 + v_{\rm eff}(\bm r, t)] \chi_\nu(\bm r) d\bm
r$ in an atomic basis set $\{\chi_{\mu}(\bm r)\}$. The square
bracket on the right--hand side (RHS) of \Eq{eom-full} denotes a commutator. The
matrix element of $\bm \sigma$ is defined as $\bm \sigma_{\mu\nu}(t)
\equiv \langle a^\dag_\nu(t) a_\mu(t) \rangle$, where $a_\mu(t)$ and
$a^\dag_\nu(t)$ are Heisenberg annihilation and creation operators
for an electron occupying atomic orbitals $\mu$ and $\nu$ at time
$t$, respectively. Fourier--transformed into frequency domain while
considering linear response only, \Eq{eom-full} leads to the
conventional Casida's equation.\cite{Cas96}

\subsection{RSDM based TDDFT--NEGF--EOM approach for open systems}
\label{tddft-eom}

In 2004, Fournais~\etal have proven the real analyticity of electron
density functions of any atomic or molecular
eigenstates.\cite{fourn}
%
%
The real analyticity property leads to the holographic theorem of
\emph{time--independent} systems: The ground--state electron density
on any finite subsystem determines completely the electronic
properties of the entire system.\cite{zheng1}

As for \emph{time--dependent} systems,
%
we have proven \emph{Holographic Time--Dependent Electron
Density Theorem}:\cite{zheng2}
%
Let $v(\bm r, t)$ be the time--dependent external potential field on
a finite physical system, $\rho(\bm r, t_0)$ be its electron density
function at a given time $t_0$, and $\rho_{\rm D}(\bm r, t)$ the
electron density function within any finite subspace D. 
If $\rho(\bm r, t_0)$ is real analytic in $\bm r$--space
and $v(\bm r, t)$ real analytic in both $\bm r$ and $t$,
there is  a \emph{one--to--one} correspondence between
$\rho_{\rm D}(\bm r, t)$  and $v(\bm r, t)$;
consequently, $\rho_{\rm D}(\bm r, t)$
determines uniquely the full {\it ab initio} Hamiltonian, and thus,
all electronic properties of the entire time--dependent
physical system. In principle, all one needs to know is
the electron density in a local subsystem.

Based on the above theorem, we have developed a formally exact
TDDFT--EOM formalism for open electronic systems. We consider coherent
time--dependent quantum transport through a realistic electronic
device. The entire system consists of bulk electrodes and the device
region, where electron--electron scattering events mainly take
place. For instance, in a two--terminal setup, the device region
(D), left electrode (L), and right electrode (R) form the entire
system. The region D is thus the open electronic system of primary
interest.

Expanded in an atomic orbital basis set, the RSDM $\bm \sigma$ of
entire system can be partitioned into a number of blocks, among
which $\bm \sigma_{\alpha}$ and $\bm \sigma_{\rm D}$ are the
diagonal blocks corresponding to the electrode $\alpha$ ($\alpha =$
L or R) and the device region D, respectively; and $\bm \sigma_{\alpha
\rm D} = \bm \sigma_{\rm D \alpha}^\dag$ representing the
off--diagonal block between the electrode $\alpha$ and the region D.
The KS Fock matrix $\bm h$ can be partitioned in a similar fashion.

To obtain the reduced electronic dynamics of the device, we focus on
the EOM of $\bm \sigma\dd$:
\be
   i\dot{\bm \sigma\dd} = \left[ \bm h\dd, \bm \sigma\dd \right] -
   i\sum_{\alpha} \bm Q_\alpha(t). \label{eom-sigma-d}
\ee
Here, $\bm Q_\alpha$ is the dissipation term due to the
electrode $\alpha$:
\be
   \bm Q_\alpha(t) = i\left( \bm h_{\rm D \alpha} \bm\sigma_{\alpha \rm D}
   - \bm\sigma_{\rm D\alpha} \bm h_{\alpha\rm D} \right).
\ee
At first glance \Eq{eom-sigma-d} seems not closed. However, based on
\emph{Holographic Time--Dependent Electron Density Theorem}, all
physical quantities are explicit or implicit functionals of the
electron density of the system D, $\rho\dd(\bm r, t)$, and so is
$\bm Q_\alpha(t)$. Therefore, \Eq{eom-sigma-d} can be cast into a
formally closed form:
\be
  i\dot{\bm\sigma\dd} = \left[ \bm h\dd\left[t;\rho\dd(\bm r,t)\right], \bm\sigma\dd \right]
  - i\sum_\alpha \bm Q_\alpha\left[t;\rho\dd(\bm r,t)\right],
  \label{eom0}
\ee
by noting that $\rho\dd(\bm r,t)$ is the diagonal content of
$\bm\sigma\dd(t)$ in $\bm r$--space.
Equation~(\ref{eom0}) is the heart of RSDM based TDDFT--EOM approach.
The transient electric current through the interface $S_\alpha$ (the
cross section separating region D from the electrode $\alpha$) can
be evaluated via
\be \label{curr-time}
   J_\alpha(t) = -\int_{\alpha} d\bm r \frac{\partial}{\partial t}
   \rho(\bm r,t) = - {\rm tr} \left[ \bm Q_\alpha(t)  \right].
\ee

The challenge now is to construct a formally exact and numerically
accessible expression for $\bm Q_\alpha[t;\rho\dd(\bm r,t)]$. \cite{Jin08234703}
Based on the Keldysh NEGF formalism, we have \cite{zheng2}
\begin{align}
 \bm Q_\alpha(t) = &- \int_{- \infty}^{t} d\tau [ \bm G_{\dd}^{r} (t,\tau) \bm\Sigma_{\alpha}^{<} (\tau,t) \nl
                  &+ \bm G_{\dd}^{<} (t,\tau) \bm\Sigma_{\alpha}^{a} (\tau,t) ] + {\rm
                  H.c.}
  \label{Q-term}
\end{align}
Approximate schemes based on NEGF have been proposed in
Ref.~\onlinecite{zheng2}.

The exact HEOM--QDT provides another viable option for the
evaluation of $\bm Q_\alpha[t;\rho\dd(\bm r,t)]$. The resulting
TDDFT--HEOM approach is expected to improve systematically the
accuracy of first--principles simulation on time--dependent quantum
transport through realistic electronic devices.

\subsection{Hierarchical TDDFT for open electronic systems}
\label{hierar}

Differing from the conventional QDTs which assume weak
device--electrode couplings, the HEOM--QDT formalism is considered
as nonperturbative.\cite{yan1} We consider a reference
noninteracting electronic system within the TDKS framework:
 \be \hat{H}_{\rm T} = \hat{H}_{\rm D} + \sum_\alpha (\hat{H}_\alpha +
    \hat{H}_{\alpha {\rm D}}).
 \ee
Here, $\hat{H}_{\rm D}$ is the KS Hamiltonian matrix for the device
D, $\hat{H}_\alpha$ is the KS Hamiltonian matrix for the electrode
$\alpha$, and $\hat{H}_{\alpha {\rm D}}$ is the KS Hamiltonian
matrix for the interaction between D and $\alpha$. They are
expressed as follows in second--quantization,
\begin{align}
   \hat{H}_{\rm D} &= \sum_{\mu\nu \in {\rm D}} \bm h_{\mu\nu} a^\dagger_\mu a_\nu, \nl
   \hat{H}_\alpha  &= \sum_{k\in\alpha} \epsilon_{\alpha k}\, d^\dagger_{\alpha k}
   d_{\alpha k}, \nl
   \hat{H}_{\alpha {\rm D}} &= \sum_{\mu\in {\rm D}}\sum_{k\in\alpha}
   \bm t_{\alpha k \mu} d^\dagger_{\alpha k}a_{\mu} + {\rm H.c.}
\end{align}
Here, $a^\dagger_\mu$ and $a_\nu$ are creation and annihilation
operators for the reference electrons in the device D, respectively.
$d^\dagger_{\alpha k}$ and $d_{\alpha k}$ are the creation and
annihilation operators for the reference electrons in the electrode
$\alpha$, respectively. $\bm h_{\mu\nu} = \langle \mu | \hat{h}(\bm
r, t) | \nu \rangle = \int \chi_\mu(\bm r) [-\frac{1}{2}\nabla^2 +
v_{\rm eff}(\bm r, t)]\chi_\nu(\bm r) d\bm r$, $\eps_{\alpha k} =
\langle k_\alpha |\hat{h}(\bm r, t) | k_\alpha \rangle$, and $\bm
t_{\alpha k\mu} = \langle k_\alpha |\hat{h}(\bm r, t) | \mu
\rangle$.
The EOM for the density matrix of the entire system (system plus
electrodes), $\hat{\rho}_{\rm T}$, is
\begin{align}
   \dot{\hat{\rho}}_{\rm T} &= -i[\hat{H}_{\rm T}, \hat{\rho}_{\rm T}] \nl
   &= -i\big[\hat{H}_{\rm D} + \sum_\alpha \hat{H}_\alpha, \hat{\rho}_{\rm
   T}\big] \nl
   &\quad - i\sum_{\alpha}\sum_{\mu\in{\rm D}} \big[f^\dagger_{\alpha\mu}a_\mu
   +a^\dagger_\mu f_{\alpha\mu}, \hat{\rho}_{\rm T} \big],
   \label{dot_hat_rho_T}
\end{align}
with $f_{\alpha\mu} = \sum_k \bm t_{\alpha k \mu}^\ast d_{\alpha k}$
and $f^\dagger_{\alpha\mu} = \sum_k \bm t_{\alpha k
\mu}d^\dagger_{\alpha k}$. \vspace{0.5cm}

The TDDFT--EOM for the reduced system, \Eq{eom0}, can be recovered.
With the equality  that $\bm \sigma_{\mu\nu}(t)={\rm tr}_{\rm T} [
a^+_\nu a_\mu \hat{\rho}_{\rm T} (t)]$, \Eq{dot_hat_rho_T} leads to
(\emph{cf.} \App{ab_deriv_HEOM})
 \begin{align}
   i\dot{\bm\sigma\dd}
   &= \left[\bm h\dd, \bm\sigma\dd \right] - \sum_\alpha
   \left[ \bm\varphi_\alpha(t) -
   \bm\varphi_\alpha^\dag(t)\right],
   \label{eom-sigma-heom}
 \end{align}
where $\bm\varphi_{\alpha ,\mu\nu}(t) \equiv {\rm tr}_{\rm T}
[ a^+_\nu f_{\alpha\mu} \hat{\rho}_{\rm T} (t) ]$
and $\bm\varphi^\dagger_{\alpha ,\mu\nu}(t) \equiv {\rm tr}_{\rm T}
[ f^{\dagger}_{\alpha\nu} a_\mu \hat{\rho}_{\rm T} (t) ]$.

By comparing \Eqs{eom0} and (\ref{eom-sigma-heom}), it is apparent
the dissipation functional $\bm Q_\alpha[\rho_{\rm D}(\bm r, t)]$ is
directly associated with the first--tier auxiliary RSDM,
$\bm\varphi_\alpha(t)$, as follows,
\begin{align}
   \bm Q_\alpha(t) &= -i \left[ \bm\varphi_\alpha(t) -
   \bm\varphi_\alpha^\dag(t) \right] \nl
   &= - i\int d\eps \left[ \bm\varphi_\alpha(\eps,t) -
   \bm\varphi_\alpha^\dag(\eps,t)\right], \label{q-term-heom}
\end{align}
where $\bm\varphi_\alpha(t)=\int d\eps\; \bm\varphi_\alpha(\eps,t)$
and $\bm\varphi_\alpha^\dag(t)=\int
d\eps\;\bm\varphi_\alpha^\dag(\eps,t)$.
\vspace{0.2cm}

A key feature of the exact HEOM--QDT is that for noninteracting
systems, such as the TDKS reference system in the present case, the
hierarchy terminates exactly at the second tier without any
approximation.\cite{Jin08234703} The corresponding HEOM for the KS
RSDM and its auxiliary counterparts have been derived in
Ref.~\onlinecite{Jin08234703} as follows,
 \begin{align}
   i\dot{\bm\varphi_\alpha}(\eps,t) &= \left[ \bm h\dd(t) - \eps - \Delta_\alpha(t) \right]
   \bm\varphi_\alpha(\eps,t) \nl &\quad+ \left[f_\alpha(\eps) - \bm \sigma\dd
   \right]
    \bm\Lambda_\alpha(\eps) \nl &\quad + \sum_{\alpha'}
   \int d\eps' \bm\varphi_{\alpha,\alpha'}(\eps,\eps',t), \label{heom-1st} \\
   i\dot{\bm\varphi}_{\alpha,\alpha'} (\eps, \eps^\prime,t)&= -\left[
   \eps + \Delta_\alpha(t) - \eps' - \Delta_{\alpha'}(t) \right]
   \bm\varphi_{\alpha,\alpha'} \nl
   &\quad + \bm\Lambda_{\alpha'}(\eps') \bm \varphi_\alpha(\eps,t)
   - \bm \varphi^\dag_{\alpha'}(\eps',t) \bm\Lambda_{\alpha}(\eps).
   \label{heom-2nd}
 \end{align}
Here, $f_\alpha(\eps) = 1/[e^{\beta(\eps-\mu_\alpha)} + 1]$ is the
Fermi distribution function for the electrode $\alpha$, with $\beta$
being the inverse temperature and $\mu_\alpha$ the equilibrium Fermi
energy; $\Delta_\alpha(t) = -V_\alpha(t)$ is the energy shift for
all single--electron levels in electrode $\alpha$ due to the
time--dependent voltage applied on $\alpha$; $\bm\Lambda_{\alpha,
\mu\nu}(\eps) \equiv \sum_{k\in\alpha} \delta(\eps-\eps_k) \bm
t_{\alpha k\mu}^\ast \bm t_{\alpha k\nu}$, is the device--electrode
coupling matrix.

Equations~(\ref{eom-sigma-heom}), (\ref{heom-1st}) and
(\ref{heom-2nd}) complete the TDDFT--HEOM formalism which is
formally exact and closed. The basic variables involved are the
reduced single--electron quantities $\{\bm\sigma\dd (t),
\bm\varphi_\alpha(\eps,t),
\bm\varphi_{\alpha,\alpha'}(\eps,\eps',t)\}$. \vspace{0.2cm}

\subsection{TDDFT--NEGF--HEOM for quantum transport}
\label{T--N--H-q-t}

We now make connection between the TDDFT--HEOM formalism to the RSDM
based TDDFT--NEGF--EOM formalism.
%
As shown in \App{ab_deriv_HEOM}, the auxiliary RSDMs
$\{\bm\varphi_\alpha(\eps,t),
\bm\varphi_{\alpha,\alpha'}(\eps,\eps',t)\}$ can be expressed in
terms of NEGF quantities as follows,\cite{croy2}
\begin{widetext}
 \begin{align}
   \bm \varphi_\alpha(\eps,t) &= i \int_{-\infty}^t\!\! d\tau
   \Big[ \bm G\dd^<(t,\tau)\, \bm \Sigma^>_\alpha(\tau,t;\eps)
  - \bm G\dd^>(t,\tau)\, \bm \Sigma^<_\alpha(\tau,t;\eps)
   \Big], \label{phi-1st}  \\
   \bm \varphi_{\alpha,\alpha'}(\eps,\eps',t)
   &= i\int_{-\infty}^t\!\! dt_1
   \int_{-\infty}^t\!\! dt_2 \,
   \Big[ \bm \Sigma_{\alpha'}(t,t_1;\eps')\,
  \bm G\dd(t_1,t_2)\, \bm \Sigma_{\alpha}(t_2,t;\eps) \Big]^{<} \\
   &= i\int_{-\infty}^t\!\! dt_1
   \int_{-\infty}^t\!\! dt_2 \, \Big\{
   \Big[ \bm \Sigma^<_{\alpha'}(t,t_1;\eps')\,
  \bm G\dd^a(t_1,t_2) + \bm \Sigma^r_{\alpha'}(t,t_1;\eps')\,
 \bm G\dd^<(t_1,t_2) \Big]\, \bm
   \Sigma_\alpha^>(t_2,t;\eps)
\nl&\qquad\qquad\qquad\qquad
  - \Big[ \bm \Sigma^r_{\alpha'}(t,t_1;\eps') \, \bm
   G\dd^>(t_1,t_2)
  + \bm \Sigma^>_{\alpha'}(t,t_1;\eps')\,  \bm
   G\dd^a(t_1,t_2) \Big]
 \bm \Sigma_\alpha^<(t_2,t;\eps) \Big\} .
   \label{phi-2nd}
 \end{align}
\end{widetext}
Here, $\bm \Sigma^x_\alpha(t,\tau;\eps)$ are frequency--dispersed
self--energies ($x=r, a, <$ and $>$), which normalize to normal
self--energies as $\int \bm \Sigma^x_\alpha(t,\tau;\eps)\, d\eps =
\bm \Sigma^x_\alpha(t,\tau)$. It is straightforward to derive from
\App{sstate} that
 \begin{align}
  \bm \Sigma^x_\alpha(t,\tau;\eps) &\equiv \exp\left[-i\int_\tau^t
  \Delta_\alpha(\zeta)\, d\zeta \right] e^{-i\eps(t-\tau)}
  \tilde{\bm \Sigma}^x_\alpha(\eps) \nl
  &= \exp\left[-i\int_\tau^t\Delta_\alpha(\zeta)\, d\zeta \right]
  \tilde{\bm\Sigma}^x_\alpha(t-\tau;\eps).
 \end{align}
This is formally analogous to normal self--energies with $\tilde{\bm
\Sigma}^x_\alpha$ being the ground/equilibrium--state quantities;
see \Eq{self-x}.
The following EOM thus applies:
 \begin{align}
   \frac{\partial}{\partial t} \bm \Sigma^x_\alpha(t,\tau;\eps)
   &= -i \left[ \Delta_\alpha(t) + \eps \right] \bm
   \Sigma^x_\alpha(t,\tau;\eps), \nl
   \frac{\partial}{\partial \tau} \bm \Sigma^x_\alpha(t,\tau;\eps)
   &= \;\;\, i \left[ \Delta_\alpha(\tau) + \eps \right] \bm
   \Sigma^x_\alpha(t,\tau;\eps).
   \label{EOM-Sigma}
 \end{align}
Equations (\ref{eom-sigma-heom}) and (\ref{heom-1st}--\ref{phi-2nd})
are the central equations of the TDDFT--NEGF--HEOM formalism.

The stationary solution of TDDFT--NEGF--HEOM is related to the
Landauer--B\"{u}ttiker formula; see
\App{sstate} for the details and remarks on this issue.

\section{Hierarchical equations of motion scheme for calculation of
dissipation functional} \label{q-exact}

\subsection{Frequency dispersion scheme for TDDFT--HEOM} \label{zero-temp}

In order to numerically solve the TDDFT--HEOM for $\bm Q_\alpha(t)$,
the integration over continuous function needs to be transformed
into summation of discrete terms, \emph{i.e.}, $\int g(\eps)\,d\eps
\rightarrow \sum_{k=1}^{N_k} w_k\,g(\eps_k)$. Here, $\eps_k$ and
$w_k$ are the $k\,$th \emph{real} frequency grid and its associated
weight ($0 \le w_k \le 1$ and $\sum_{k=1}^{N_k} w_k = 1$),
respectively. We have thus
 \be
   \bm Q_\alpha(t) = - i \sum_{k=1}^{N_k} w_k \left[ \bm\varphi_{\alpha k}(t)
   - \bm\varphi_{\alpha k}^\dag(t) \right].
 \ee
Equations~(\ref{heom-1st}) and (\ref{heom-2nd}) are recast into
 \begin{align}
    i\dot{\bm\varphi}_{\alpha k} &= \left[\bm h\dd(t) - \eps_k - \Delta_\alpha(t) \right]
    \bm\varphi_{\alpha k}(t) + \left[ f_\alpha(\eps_k) - \bm\sigma\dd
    \right]
 \nl
    &\quad \times \bm\Lambda_\alpha(\eps_k) +
    \sum_{\alpha'}\sum_{k'=1}^{N_k} w_{k'} \bm\varphi_{\alpha k,\alpha'
    k'}(t), \label{heom-1a}\\
    i\dot{\bm\varphi}_{\alpha k, \alpha' k'} &= -\left[ \eps_k +
    \Delta_\alpha(t) - \eps_{k'} - \Delta_{\alpha'}(t) \right]
    \bm\varphi_{\alpha k, \alpha' k'}(t)
 \nl
    &\quad + \bm\Lambda_{\alpha'}(\eps_{k'}) \bm\varphi_{\alpha
    k}(t) - \bm\varphi_{\alpha' k'}^\dag(t)
    \bm\Lambda_\alpha(\eps_k), \label{heom-2a}
 \end{align}
Hereafter, we adopt abbreviations $\bm\varphi_{\alpha k}(t) \equiv
\bm\varphi_{\alpha}(\eps_k,t)$ and $\bm \varphi_{\alpha k, \alpha'
k'}(t) \equiv \bm \varphi_{\alpha,\alpha'}(\eps_k, \eps_{k'}, t)$.
In practice, application of quadrature rules, such as
Gauss--Legendre quadrature, reduces significantly the number of
frequency grids compared to an equidistant sampling. This
frequency--dispersed TDDFT--HEOM scheme applies to both zero and finite
temperatures. Next we propose several efficient schemes particularly
designed for finite temperature cases.

\subsection{Finite temperature schemes for TDDFT--NEGF--HEOM} \label{finite-temp}

The ground/equilibrium--state self--energies
$\tilde{\bm\Sigma}^{x}_{\alpha}(t)$ are usually exponentially
decaying functions of time. Therefore, the following exponential
expansion is attainable for $\tilde{\bm\Sigma}^{<,>}_\alpha(\tau-t)$
at $t \geq \tau$:
 \begin{align}
   \tilde{\bm\Sigma}^{x}_{\alpha}(\tau-t) &\simeq \sum_{k=1}^{N_k}
   \bm A_{\alpha k}^{x} \exp\left[ -\gamma_{\alpha k}
   (t-\tau)\right] \nl
   &\equiv \sum_{k=1}^{N_k} \tilde{\bm\Sigma}^{x}_{\alpha
   k}(\tau-t) \label{sigma-sd}
 \end{align}
for $x = <$ and $>$. $\gamma_{\alpha k}$ are $c$--numbers with ${\rm
Re}(\gamma_{\alpha k}) > 0$. Moreover, the nonequilibrium
self--energies $\bm\Sigma^x_{\alpha k}(\tau, t) \equiv e^{-i
\int^\tau_t \Delta_\alpha(\zeta)d\zeta} \tilde{\bm\Sigma}^x_{\alpha
k}(\tau-t)$. Equation~(\ref{q-term-heom}) becomes
 \be
  \bm Q_\alpha(t) = - i \sum_{k=1}^{N_k} \left[ \bm\varphi_{\alpha k}(t)
   - \bm\varphi_{\alpha k}^\dag(t) \right].
  \label{Q-alpha}
 \ee
Here, $\bm\varphi_{\alpha k}(t)$ is given by the RHS of \Eq{phi-1st}
with $\bm\Sigma^x_\alpha(\tau, t;\eps)$ replaced by
$\bm\Sigma^x_{\alpha k}(\tau, t)$. The second--tier auxiliary RSDMs,
$\bm\varphi_{\alpha k, \alpha' k'}(t)$, are formally obtained by
substituting $\bm\Sigma^x_\alpha(t_2, t;\eps)$ with
$\bm\Sigma^x_{\alpha k}(t_2, t)$ on the RHS of \Eq{phi-2nd}. EOM for
$\bm\varphi_{\alpha k}$ and $\bm\varphi_{\alpha k, \alpha' k'}(t)$
now read:
 \begin{align}
    i\dot{\bm\varphi}_{\alpha k} &= \left[\bm h\dd(t) - i \gamma_{\alpha
    k} - \Delta_\alpha(t) \right] \bm\varphi_{\alpha k}(t) \nl
    &\quad + i \left[ \bm\sigma\dd(t) \bm A^>_{\alpha k} +
    \bar{\bm\sigma}\dd(t) \bm A^<_{\alpha k} \right] \nl
    &\quad + \sum_{\alpha'}\sum_{k'=1}^{N_k} \bm\varphi_{\alpha k,\alpha'
    k'}(t), \label{i-varphidot-alpha-k} \\
    i\dot{\bm\varphi}_{\alpha k, \alpha' k'} &= -\left[ i\gamma_{\alpha k} +
    \Delta_\alpha(t) - i\gamma_{\alpha' k'} - \Delta_{\alpha'}(t) \right]
    \bm\varphi_{\alpha k, \alpha' k'} \nl
    &\quad + i\left(\bm A^>_{\alpha' k'} - \bm A^<_{\alpha' k'} \right) \bm\varphi_{\alpha
    k}(t) \nl
    &\quad - i\, \bm\varphi_{\alpha' k'}^\dag(t) \left(
    \bm A^>_{\alpha k} - \bm A^<_{\alpha k} \right) \label{i-varphidot-alpha-k-alphaprime-kprime}.
 \end{align}
Here, $\bar{\bm\sigma}\dd \equiv 1 - \bm\sigma\dd$ is the KS reduced
single--hole density matrix. \Eqs{eom-sigma-heom},
(\ref{i-varphidot-alpha-k}) and
(\ref{i-varphidot-alpha-k-alphaprime-kprime}) constitute a
reformulation of TDDFT--NEGF--HEOM formalism which is suitable for
the numerical solution.

Apparently, the computational cost for solving the TDDFT--NEGF--HEOM is
determined by the number of exponential functions $N_k$ used to
expand the lesser and greater self--energies. Next we propose three
decomposition schemes to expand the self--energies as in
Eq.~(\ref{sigma-sd}).

%
%
\subsubsection{Matsubara expansion decomposition scheme} \label{md-scheme}
The ground/equilibrium--state lesser and greater self--energies can
be evaluated via
 \begin{align}
   \tilde{\bm\Sigma}_\alpha^x(\tau-t) &= \varsigma_x i \int d\eps\,
   e^{i\eps (t-\tau)} f_\alpha^{\varsigma_x}(\eps) \bm\Lambda_\alpha(\eps)
   \label{self-fdt}
 \end{align}
for $t \geq \tau$. Here, $x = <$ or $>$, $\varsigma_< = +$ and
$\varsigma_> = -$, and $f^-_\alpha(\eps) \equiv f_\alpha(\eps)$ and
$f^+_\alpha(\eps) \equiv 1 - f_\alpha(\eps)$.
Equation~(\ref{self-fdt}) states the fluctuation--dissipation
theorem for electrode correlation functions.
If the linewidth matrix $\bm\Lambda_\alpha(\eps)$ can be approximated
via a multi--Lorentzian expansion as follows,
\be
   \bm\Lambda_\alpha(\eps) = \sum_{d=1}^{N_d}
   \frac{\eta_d}{(\eps-\Omega_d)^2 +
   W_d^2}\,\bar{\bm\Lambda}_{\alpha d}. \label{lambda-engy}
\ee
Here, $\eta_d$, $\Omega_d$, and $W_d > 0$ are the coefficient,
center, and width of $d$th fitting Lorentzian function; and
$\bar{\bm\Lambda}_{\alpha d}$ is the corresponding
frequency--independent linewidth matrix, respectively. The RHS of
\Eq{self-fdt} is calculated via complex contour integral and residue
theorem, for which analytical continuation of
$\bm\Lambda_\alpha(\eps)$ and $f^\pm_\alpha(\eps)$ into complex plane
is needed. $\bm\Lambda_\alpha(z)$ can be defined straightforwardly by
replacing $\eps$ with $z$ on the RHS of \Eq{lambda-engy}, and
$f^\pm_\alpha(\eps)$ is continued analytically to $f^\pm_\alpha(z)$
in the same fashion.
At a finite temperature, the Matsubara expansion scheme for the
Fermi function is as follows,
 \begin{align}
   f^\varsigma_\alpha(z) &\equiv
   \frac{1}{e^{\varsigma\beta(z-\mu_\alpha)}+1} \nl
   &\simeq \frac{1}{2} - \varsigma
   \frac{1}{\beta} \sum_{p=1}^{N_p}\left( \frac{1}{z + z_{\alpha p}} +
   \frac{1}{z - z_{\alpha p}} \right). \label{fermi-approx}
 \end{align}
For the Matsubara expansion $z_{\alpha p} = \mu_\alpha + i\pi(2p-1)/\beta$.

With all the poles of $f^\varsigma_\alpha(z)$ and
$\bm\Lambda_\alpha(z)$ in the upper--half complex plane accounted
for, we arrive at the following exponential expansion of
self--energies:
\be
   \tilde{\bm\Sigma}^x_\alpha(\tau-t) \simeq
   \sum_{d=1}^{N_d} \bm A^x_{\alpha
   d}\,e^{-\gamma_{\alpha d}(t-\tau)} +
   \sum_{p=1}^{N_p} \bm B^x_{\alpha
   p}\,e^{-\check{\gamma}_{\alpha p}(t-\tau)}, \label{sigma-mats}
\ee
with
\begin{align}
   \bm A^x_{\alpha d} &= i
   \frac{\varsigma_x \pi \eta_d}{W_d}f^{\varsigma_x}_\alpha(\Omega_d + iW_d)
   \bar{\bm \Lambda}_{\alpha d}, \nl
   \gamma_{\alpha d} &= W_d - i\Omega_d, \nl
   \bm B^x_{\alpha p} &= \frac{2\pi}{\beta}
   \bm\Lambda_\alpha(z_{\alpha p}), \nl
   \check{\gamma}_{\alpha p} &= -i z_{\alpha p} = -i\mu_\alpha
   + \frac{(2p-1)\pi}{\beta}.  \label{coef-mats}
\end{align}
We then make connections between \Eq{sigma-mats} and \Eq{sigma-sd}.
The total number of exponential terms is the sum of number of
Lorentzian functions and number of Matsubara terms considered,
\emph{i.e.}, $N_k = N_d + N_p$. In principle, $N_p \rightarrow
+\infty$ is required to achieve an exact expansion for
$\tilde{\bm\Sigma}^{<,>}_\alpha(\tau-t)$. In practice, a smallest
possible $N_p$ but guaranteeing an expected accuracy is desired. The
value of $N_p$ increases rapidly as the temperature is lowered.

\subsubsection{Partial fractional decomposition scheme}
\label{pfd-scheme}

A major disadvantage of the Matsubara expansion is the poor
convergence at low temperature. Alternatively we may adopt the
partial fractional decomposition (PFD) of Fermi function.\cite{croy}
The PFD expansion is formally identical to \Eq{fermi-approx}, but
with $z_{\alpha p} = \mu_\alpha + 2\sqrt{\eps_p}/\beta$. Here,
$\eps_p$ is the $p\,$th eigenvalue of the $N_p \times N_p$ matrix
$\bm Z$ defined as follows,
\be
   \bm Z_{mn} = 2m(2m-1)\,\delta_{n,m+1} -
   2N_p(2N_p-1)\,\delta_{m,N_p}.
\ee
with the constraint that ${\rm Im}(\sqrt{\eps_p})> 0$.

With the same multi--Lorentzian expansion of \Eq{lambda-engy}, the
self--energies are decomposed into exponential functions exactly as \Eq{sigma-mats}.
The PFD scheme leads to a formally similar exponential
expansion for the self--energies, and hence for the resulting HEOM,
as compared with the widely used Matsubara expansion scheme.
However, numerical tests have confirmed that with same $N_p$, the
PFD scheme yields much more accurate approximation for
$f^\varsigma_\alpha(z)$ than the Matsubara expansion scheme.\cite{croy, Xu01} In
other words, to achieve the same level of accuracy for the
self--energies, a much smaller number of exponential terms is
required with the PFD scheme. Therefore, the PFD scheme is superior
to the conventional Matsubara expansion scheme in terms of
computational efficiency.

%
%
\subsubsection{Hybrid spectral decomposition and frequency dispersion
scheme}\label{sd-fd}

\begin{figure}[h]
\begin{center}
\includegraphics[scale=0.35]{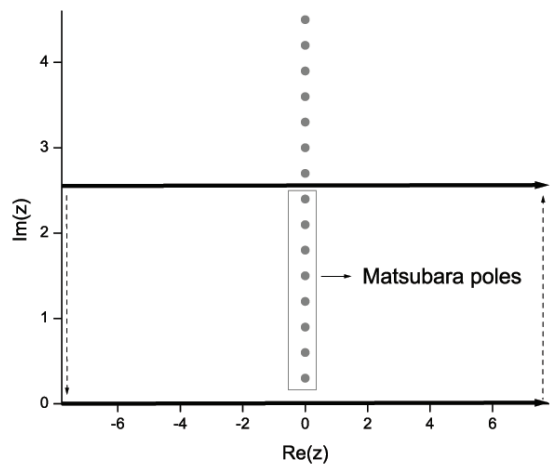}
\caption{An illustrative plot showing the contour of integration for
$f_\alpha(z)\Lambda_\alpha(z)$ with a single--Lorentzian spectral
density of $\Omega_1 = 5$. The dots equidistantly distributing on
${\rm Re}(z) = 0$ correspond to the Matsubara poles (grey dots) in
the upper complex plane. Integration along the dark solid arrow
lying right on the real axis gives the desired self--energy of
\Eq{self-fdt}. However, the resulting HEOM encounters numerical
difficulties. Due to the fact that the two dark solid arrows form a
closed loop (completed with the dark dashed arrows at infinitely
distance carrying zero values), the self--energy can be obtained
alternatively by summing up the residues at the poles within this
loop. \label{fig:contour}}
\end{center}
\end{figure}

As mentioned after \Eq{coef-mats}, in principle $N_p \rightarrow
+\infty$ is required to achieve an exact expansion for the Fermi
function, and hence for the self--energies. With a finite $N_p$, the
deviation between the RHS of \Eq{fermi-approx} and the exact
function $f^\varsigma_\alpha(z)$ results in the following
contribution to the self--energies:
\be
   \tilde{\bm\Sigma}^{x, {\rm dev}}_\alpha(\tau-t) = \varsigma_x i
   \int_{{\rm Im}(z)=y} e^{iz(t-\tau)}
   f^{\varsigma_x}_\alpha(z) \bm\Lambda_\alpha(z) \,dz
   \label{self-dev}
\ee
for $t \geq \tau$, as shown in \Fig{fig:contour}.
Different from \Eq{self-fdt} where the integration is along the real
axis, here the integration contour is a horizontal line: ${\rm
Im}(z) = y > 0$. The value of $y$ is chosen so that the first
upper--plane $N_p$ Matsubara poles reside in the interstitial region
bounded by ${\rm Im}(z) = 0$ and ${\rm Im}(z) = y$, while all the
other poles are located outside this area, \emph{i.e.},
$(2N_p-1)\pi/\beta < y < (2N_p+1)\pi/\beta$.

As long as the modulus
of $\bm\Lambda_\alpha(z)$ remains sufficiently small along the line
${\rm Im}(z) = y$, \Eq{self-dev} presents a complementary
contribution to the RHS of \Eq{sigma-mats}. The integration in
\Eq{self-dev} is transformed into summation of discrete terms,
following the frequency dispersion scheme introduced in
\Sec{zero-temp}, \emph{i.e.}, $\int_{{\rm Im}(z)=y} g(z)\,dz
\rightarrow \sum_{q=1}^{N_q} w_q\, g(\eps_q + iy)$. We have thus
\be
   \tilde{\bm\Sigma}^{x, {\rm dev}}_\alpha(\tau-t) \simeq
   \sum_{q=1}^{N_q} \bm K^x_{\alpha q} e^{-\hat{\gamma}_{\alpha
   q}(t-\tau)}, \label{sigma-dev}
\ee
with
\begin{align}
   \bm K^x_{\alpha q} &= \varsigma_x i\, w_q
   f^{\varsigma_x}_\alpha(\eps_q+iy) \bm\Lambda_\alpha(\eps_q + iy),
   \nl
   \hat{\gamma}_{\alpha q} &= - i(\eps_q + iy) = y - i\eps_q.
\end{align}
Therefore, overall $\tilde{\bm\Sigma}^x_\alpha(\tau-t)$ is given by
combining the RHS of \Eqs{sigma-mats} and (\ref{sigma-dev}). The
total number of exponential functions is $N_k = N_d + N_p + N_q$.
In practice, to make use of \Eq{sigma-dev}, one needs to assign an
appropriate number of Matsubara terms, $N_p$. On one hand, a smaller
$N_p$ leads to fewer unknown variables of HEOM. On the other hand, a
too small $N_p$ (and hence a too small $y$) would cause severe
convergence problem when solving the stationary solutions of HEOM.
Therefore, the value of $N_p$ should be chosen carefully for the
balance between computational cost and numerical difficulty. Once
the $N_p$ and $y$ are settled, efficient quadratures can then be
adopted to optimize the frequency dispersion (the set $\{w_q, \eps_q
\}$ with $q = 1,\ldots,N_q$) for \Eq{self-dev}. Finally, all the
$N_d$ poles of $\bm\Lambda_\alpha(z)$ inside the region $0 < {\rm
Im}(z) < y$ are collected.

It is important to note the value of $N_d$ depends on the
mathematical form of basis functions, based on which
$\bm\Lambda_\alpha(\eps)$ is expanded. The widely used
multi--Lorentzian expansion has been adopted in \Sec{md-scheme}; see
\Eq{lambda-engy}. Each Lorentzian function gives a pole in the upper
complex plane. One of the advantages of using a Lorentzian function
$g_{\rm L}(\eps)$ is that the magnitude of its analytically
continued counterpart, $\vert g_{\rm L}(z)\vert$, decays smoothly as
${\rm Im}(z)$ increases. Therefore, the deviation term,
$\tilde{\bm\Sigma}^{x,{\rm dev}}_\alpha(\tau-t)$ of \Eq{self-dev},
becomes consistently less significant as more Matsubara poles are
considered explicitly.
Alternative basis functions can also be used. For instance, a
Gaussian function is analytic, and does not have any pole on the
entire complex plane. Therefore, for $\bm\Lambda_\alpha(\eps)$ fitted
by a number of Gaussian functions, $N_d = 0$, which reduces the
number of unknown variables of HEOM. However, care must be taken
when treating the ``deviation'' contribution, as some of the $\bm
K^x_{\alpha q}$ may assume large values, due to the nontrivial value
of complex Gaussian function at certain $z = \eps_q + iy$.

Numerical tests on model systems\cite{Zhe09164708} have confirmed that as the
temperature lowers, the total number of exponential functions needed
to accurately resolve the self--energies in the hybrid scheme grows
much slower than that in the Matsubara expansion scheme.

%
%
%
\section{Approximate schemes for calculation of dissipation functional} \label{q-approx}

\subsection{Schemes based on wide--band limit approximation}
\label{wbl-adiabatic}

Even with the sophisticated hybrid scheme for decomposition of
self--energies, the exact HEOM approach for the reduced electronic
dynamics can still be expensive. Efficient approximate schemes are
thus required. For clarity, we will omit the spin index in the
following derivations of approximate schemes.

\subsubsection{TDDFT--NEGF--HEOM formalism under wide--band limit approximation}
\label{wbl}

The wide--band limit (WBL) approximation involves the following assumptions for the
electrodes: (i) their bandwidths are assumed to be infinitely large;
and (ii) their linewidths are assumed to be energy--independent,
\emph{i.e.}, $\bm\Lambda_\alpha(\eps) \approx \bm\Lambda_\alpha$.
Denote $\tlam_\alpha \equiv \pi\bm\Lambda_\alpha$.
From the definition of self--energies and the expansion of the Fermi function
[\Eq{fermi-approx} with $N_p \rightarrow +\infty$], one obtains for $t > \tau$
\begin{align}
\bm\Sigma^{<,>}_\alpha(\tau,t) &\equiv \pm i\, e^{i\int_\tau^t dt_1 \Delta_\alpha(t_1)}
   \int d\eps f^\pm_\alpha(\eps)\, e^{i\eps(t-\tau)} \bm \Lambda_\alpha(\eps)  \nl
 &= \pm i \delta(t-\tau) \tlam_\alpha \nl
 &\quad + \frac{2}{\beta} \sum_{k=1}
 e^{i \int_\tau^t dt_1 z_{\alpha k}(t_1)} \tlam_\alpha,
\end{align}
where $z_{\alpha k}(t) = z_{\alpha k} + \Delta_\alpha(t)$. The expansion of
Fermi function thus leads to a sum of exponentials
for the self--energies.
We introduce \emph{auxiliary self--energies}
$\hat{\bm\Sigma}_{\alpha k}$ to account for the exponentials,
\emph{i.e.},
\begin{align}
   \bm\Sigma^{<,>}_\alpha(\tau,t) &= \pm i \delta(t-\tau) \tlam_\alpha
   + \sum_k \hat{\bm\Sigma}_{\alpha k} (\tau,t),
   \label{self-aux-self} \\
   \hat{\bm\Sigma}_{\alpha k} (\tau,t) &= \frac{2}{\beta}\,
   e^{i\int_\tau^t dt_1 z_{\alpha k} (t_1)} \tlam_\alpha.
   \label{aux-self}
\end{align}
It is implied that $\hat{\bm\Sigma}_{\alpha k} (t, t^+) =
\frac{2}{\beta} \tlam_\alpha$. Next, we insert relation
\Eq{self-aux-self} into \Eq{phi-1st}, and arrive at
\be
   \bm\varphi_\alpha(t) = i \left[ \bm 1- 2 \bm \sigma\dd(t)\right] \tlam_\alpha
    + \sum_{k} \hat{\bm\varphi}_{\alpha k} (t),
   \label{varphi-alpha}
\ee
with the \emph{auxiliary current matrices} being as follows,
\begin{align}
   \hat{\bm\varphi}_{\alpha k}(t) &= -i \int_{-\infty}^t d\tau \left[ \bm G\dd^< (t,\tau)
   - \bm G\dd^>(t, \tau) \right] \hat{\bm\Sigma}_{\alpha k}(\tau,t) \nl
   &= i \int_{-\infty}^t d\tau\, \bm G\dd^r(t, \tau)\,
   \hat{\bm\Sigma}_{\alpha k}(\tau,t). \label{varphi-alpha-p}
\end{align}
Under the WBL approximation, NEGF formalism gives
\begin{align}
    \bm G\dd^r(t,\tau) &= -i \vartheta(t-\tau)\, \bm W\dd^-(t)\, \bm W\dd^+(\tau), \\
    \bm W\dd^\pm(t) &= \exp_{\pm}\left\{ \mp i\int_{0}^t d t_1
    \left[ \bm h\dd(t_1) - i \tlam \right] \right\}.
\end{align}
Here, we take the time from which the voltage emerges as $t_0 = 0$,
\emph{i.e.}, $\Delta_\alpha(t\leq 0) = 0$, and $\tlam =
\sum_{\alpha}\tlam_\alpha$. Therefore, the EOM of
$\hat{\bm\varphi}_{\alpha k}(t)$ can be established readily as
follows,
\be
   \dot{\hat{\bm\varphi}}_{\alpha k}(t) = \frac{2}{\beta} \tlam_\alpha
   - i \left[\bm h\dd(t) - i\tlam - z_{\alpha k}(t)\right]
   \hat{\bm\varphi}_{\alpha k}(t). \label{eom-varphi-alpha-p}
\ee
The coupled equations of motion, \Eqs {eom-sigma-d}, (\ref{Q-alpha}), (\ref{varphi-alpha})
and (\ref{eom-varphi-alpha-p}) can be solved together for a
complete description of the nonequilibrium electronic dynamics of
the device. Equation~(\ref{eom-varphi-alpha-p}) is self--closed,
which suggests that under the WBL approximation the TDDFT--HEOM
automatically terminates at first tier, instead of second--tier
termination for the exact formalism. Thus the additional
second--tier auxiliary matrices\cite{Jin08234703} are not needed
with the WBL. Similar apporach has been adopted by Croy and
Saalmann.\cite{croy}

\subsubsection{TDDFT--NEGF--EOM formalism under adiabatic wide--band limit approximation}
\label{awbl}

We consider another approximate scheme for $\bm Q_\alpha(t)$ based
on WBL approximation for electrodes, as well as an adiabatic
approximation for memory effect. The scheme aims at simplifying the
TDDFT--NEGF formulation for $\bm Q_\alpha(t)$; see
\Eq{Q-term}.
Due to the WBL approximation, the ground/equilibrium--state
self--energies become simply
\begin{align}
   \tilde{\bm\Sigma}^a_\alpha(\tau,t) &= i \delta(t-\tau)
   \tlam_\alpha, \label{sigma-a-wbl} \\
   \tilde{\bm\Sigma}^<_\alpha(\tau,t) &= i \frac{1}{\pi} \left[ \int_{-\infty}^{+\infty}
   f_\alpha(\eps)\,e^{i\eps (t-\tau)} \,d\eps \right]
   \tlam_\alpha. \label{sigma-less-wbl}
\end{align}
By inserting \Eqs{sigma-a-wbl} and (\ref{sigma-less-wbl}) into
\Eq{Q-term}, the dissipation functional is formally simplified to
be
\begin{align}
   \bm Q^{\rm AWBL}_\alpha(t) &= \big\{ \tlam_\alpha, \bm\sigma\dd \big\}
   + \bm P_\alpha(t) + \left[ \bm P_\alpha(t) \right]^\dag,
   \label{q-term-wbl} \\
   \bm P_\alpha(t) &= - \int_{-\infty}^{+\infty} \bm
   G\dd^r(t,\tau) \bm\Sigma^<_\alpha(\tau, t)\,d\tau.
   \label{p-term-wbl}
\end{align}
Before the external voltage is applied, the entire composite system
is in its ground/equilibrium state. We have
\begin{align}
   \bm P_\alpha(t) &= -\frac{i}{\pi}\,\bm U^+_\alpha(t)
   \Big\{\int_{-\infty}^{+\infty} f_\alpha(\eps)
   \Big[\frac{1}{\eps - \bm h\dd(0) + i \tlam} \nl
   &\quad -i \int_0^t e^{-i\eps\tau}
   \bm U^-_\alpha(\tau)\,d\tau\Big] e^{i\eps t} d\eps\Big\}
   \tlam_\alpha. \label{p-term-full} \\
   \bm U^\pm_\alpha(t) &\equiv \exp_\pm \Big\{ \mp \! i \!
   \int_0^t d\tau \!\left[ \bm h\dd(\tau) -i\tlam -
   \Delta_\alpha(\tau)\right]\Big\}. \label{u-term-wbl}
\end{align}
To facilitate the calculation of $\bm P_\alpha(t)$ given by
\Eq{p-term-full}, an adiabatic approximation is adopt to evaluate
the time integral within the square bracket on the RHS. The basic
idea is to first disregard the time--dependence of $\bm h\dd(t)$ and
$\Delta_\alpha(t)$ in \Eq{u-term-wbl}, evaluate the time
integral, and then restore subsequently the time--dependence of $\bm
h\dd(t)$ and $\Delta_\alpha(t)$. The final approximated expression
for $\bm P_\alpha(t)$ is as follows,
\begin{align}
   \bm P_\alpha(t) &\simeq -\frac{i}{\pi} \bigg\{ \bm U^+_\alpha(t)
   \int_{-\infty}^{+\infty} d\eps f_\alpha(\eps)\, e^{i\eps t} \nl
   &\quad \times \left[ \frac{1}{\eps - \bm h\dd(0) + i\tlam}
   - \frac{1}{\eps - \bm h\dd(t) + i\tlam + \Delta_\alpha(t)}
   \right]\nl
   &\quad + \int_{-\infty}^{+\infty} \frac{f_\alpha(\eps)}
   {\eps - \bm h\dd(t) + i\tlam + \Delta_\alpha(t)}\, d\eps\bigg\}
   \,\tlam_\alpha. \label{p-term-adia}
\end{align}
The RHS of \Eq{p-term-adia} gives exact $\bm P_\alpha(t)$ for the
ground state at $t \leq 0$ and the steady state at $t \rightarrow
+\infty$, provided that $\Delta_\alpha(t\rightarrow +\infty)$ is a
constant. At $0<t<+\infty$, it provides an adiabatic connection
between the ground and steady states. Equation~(\ref{p-term-adia})
can be solved conveniently, along with the EOM for $\bm
U^+_\alpha(t)$ as follows,
\be
  i\,\dot{\bm U}^+_\alpha(t) = \big[\bm h\dd(t) -i\tlam - \Delta_\alpha(t)
  \big]\, \bm U^+_\alpha(t). \label{eom-u-plus}
\ee
%
%

\subsection{Scheme based on complete second--order quantum dissipation
theory} \label{sec-qdt}

As presented in \Sec{q-exact}, the TDDFT--NEGF--HEOM for KS RSDM and
associated auxiliary matrices terminate exactly at the second tier.
This amounts to an explicit treatment of device--electrode
interaction at leading $4$th--order.\cite{Jin08234703} The number of unknown matrices
at zeroth, first, and second tier is $1$, $N_\alpha N_k$, and
$N_\alpha^2 N_k^2$, respectively.
Here, $N_k$ is the number of
exponential functions used to resolve the memory contents of
self--energies or electrode correlation functions. Obviously, the
second--tier variables dominate the computational cost for solving
the TDDFT--NEGF--HEOM. Therefore, a straightforward way to reduce the
computational expense is to avoid explicit involvement of the
second--tier variables, \emph{i.e.}, to truncate the TDDFT--NEGF--HEOM at
first tier approximately.

A simple truncation scheme is to set all second--tier auxiliary
matrices to zero, \emph{i.e.}, $\bm\varphi_{\alpha k, \alpha' k'}(t)
= \bm 0$. This corresponds to the chronological ordering
prescription of $2$nd--order QDT.

An alternative $2$nd--order approximation is as follows:
\begin{align}
   \bm G\dd^<(t, \tau) &\approx i \,\bm\sigma\dd(t)\,\bm U\dd^+(t,\tau), \nl
   \bm G\dd^>(t, \tau) &\approx -i\, \bar{\bm\sigma}\dd(t)\,\bm U\dd^+(t,\tau),
\end{align}
with
\be
  \bm U\dd^\pm(t,\tau) = \exp_\pm \Big[ \mp i\int_\tau^t
  \bm h\dd(\zeta)\, d\zeta \Big]. \label{udd}
\ee
The dissipation functional becomes approximately
\be
   \bm Q_\alpha(t) \approx \left[ \bm\sigma\dd(t)\,\bm\Pi^>_\alpha(t)
    + \bar{\bm\sigma}\dd(t)\, \bm\Pi^<_\alpha(t)
    + {\rm H.c.}\right], \label{q-term-pi}
\ee
where
\begin{align}
   \bm \Pi^x_\alpha(t) &= i\int_{-\infty}^t d\tau\,
   \bm U\dd^+(t,\tau)\,\bm\Sigma^x_\alpha(\tau, t). \label{pi-term}
\end{align}
Following \Sec{q-exact}, we decompose self--energies as linear
combinations of exponential functions; see \Eq{sigma-sd}. This leads
to
\begin{align}
    \bm\Pi^x_\alpha(t) &= \sum_{k=1}^{N_k} \bm\Pi^x_{\alpha k}(t),
    \nl
    \bm\Pi^x_{\alpha k}(t) &= i\int_{-\infty}^t d\tau\, \bm
    U\dd^+(t,\tau)\,\tilde{\bm\Sigma}^x_{\alpha k}(\tau -t) \nl
    &\quad \times e^{i\int_\tau^t \Delta_\alpha(\zeta) d\zeta}.
\end{align}
The EOM for $\bm\Pi^x_{\alpha k}(t)$ is thus
\be
   i\,\dot{\bm\Pi}^x_{\alpha k}(t) = \left[ \bm h\dd(t) -i\gamma_{\alpha k}
   -\Delta_\alpha(t) \right] \bm\Pi^x_{\alpha k}(t) - \bm A^x_{\alpha
   k}, \label{eom-pi-k}
\ee
with $\bm A^x_{\alpha k}$ and $\gamma_{\alpha k}$ referring to
\Eq{sigma-sd}.

The resulting complete $2$nd--order QDT approach thus involves the
coupled EOM for $\bm\sigma\dd(t)$ and $\{\bm\Pi^x_{\alpha k}(t)\}$,
with the total number of unknown matrices being $N = 2N_\alpha N_k +
1$, much fewer than $N_\alpha^2 N_k^2$ alone. This approach can be
improved further, by formal inclusion of higher--order
device--electrode interaction into the reduced system propagator
$\bm U\dd^+(t,\tau)$. This can be achieved by attaching self--energy
to $\bm h\dd(t)$ on the RHS of \Eq{udd}.

%
%
\section{Numerical results} \label{results}

\begin{figure}
\begin{center}
\includegraphics[scale=0.30]{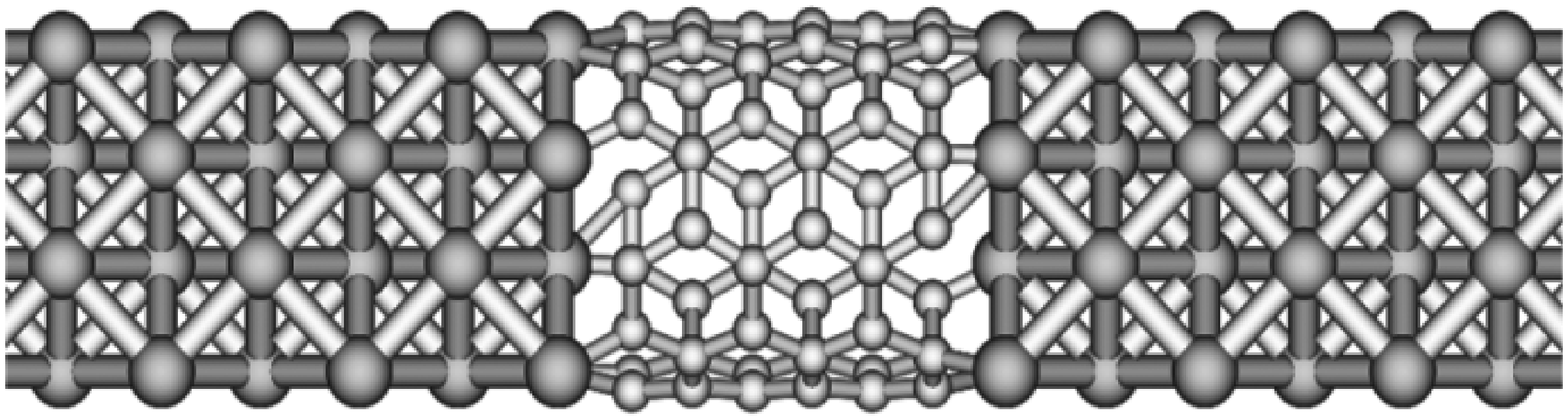}
\caption{The ball--and--stick representation of the system of
interest which is a carbon nanotube (5,5) welded to aluminium
electrodes. There are 60 carbon atoms for the carbon nanotube in
this case.}
\label{fig:al-cnt55-al}
\end{center}
\end{figure}

As a test for our practical schemes, we have performed the numerical
calculations employing the AWBL approximation
as outlined in Sec. IV A.2.
LODESTAR \cite{ldstar} was used to carried out the calculations. The system of
interest is a (5,5) carbon nanotube which is covalently bonded
between two aluminum electrodes, as shown in \Fig{fig:al-cnt55-al}.
In the simulation box, we include a finite carbon nanotube and 32
aluminum atoms for each electrode explicitly. The minimum basis set
STO--3G is adopted in the calculations. We simulate the
time--dependent electric current through the left (right) electrode
by taking the trace of the corresponding dissipative term $\bm
Q_{\rm L}$ ($\bm Q_{\rm R}$); see \Eq{curr-time}.

\begin{figure}
\begin{center}
\includegraphics[scale=0.32]{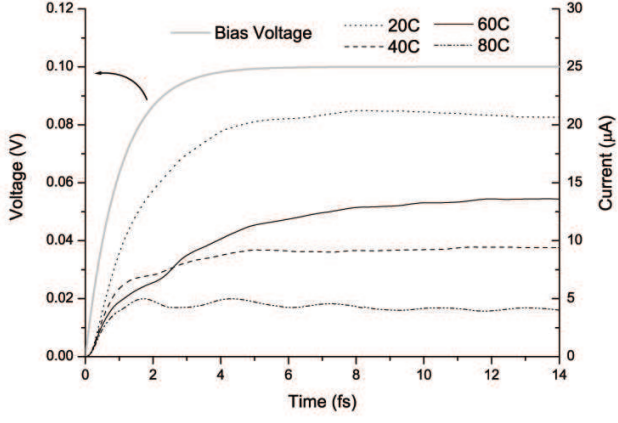}
\caption{The grey solid line represents the bias voltage applied on
the systems. The bias voltage is switched on exponentially, $V_b =
V_0 (1 - e^{-t/a})$ with $V_0 = 0.1$ V and time constant $a = 1$ fs.
The transient currents for the systems with 20, 40, 60, 80 carbon
atoms are shown dark dotted line, dashed line,
solid line and dash-dot-dot line
respectively.}
\label{fig:size-curr}
\end{center}
\end{figure}

Figure~\ref{fig:size-curr} shows the currents versus time for
different numbers of carbon atoms, \emph{i.e.} 20, 40, 60 and 80.
The bias voltage $V_b$ is switched on exponentially at $t = 0$, and
as shown in \Fig{fig:size-curr}. We observe that the currents
reach their steady states in about 8 to 12 fs for different systems. Also,
the value of the steady state current does not increase proportionally
with the size of the carbon nanotube, as that of classical case. We
note from \Fig{fig:size-curr} that the steady state currents for the
systems with 20, 40, 60 and 80 carbon atoms are about 21, 9, 14 and 4 $\mu$A, respectively.
This shows that the quantum finite size effect plays an
important role at such small scales. We point out that all the above
calculations are performed on single--CPU desktop personal computer,
and the memory is 2 gigabytes.
%

\section{Concluding remarks} \label{summary}

We have proposed a first--principles TDDFT--NEGF--HEOM method for
transient quantum transport through realistic electronic devices and
structures. The TDDFT--NEGF--HEOM formalism is in principles exact, and
formally equivalent to the TDDFT--NEGF--EOM formalism proposed
previously. In practice, it involves hierarchical equations for KS
RSDM of reduced system and associated auxiliary quantities.



By resolving memory contents of self--energies or electrode
correlation functions, we construct QDT--HEOM in the KS reduced
single--electron space. The resulting TDDFT--HEOM exactly terminate at the
second--tier. TDDFT--NEGF--HEOM formalism combines TDDFT--NEGF--EOM and
TDDFT--HEOM, and its
computational cost is determined by the number of
exponential functions used to expand the self--energies. Various
decomposition schemes are presented.

To reduce computational cost further, we have also devised approximate
schemes for TDDFT--NEGF--HEOM. One is based on the WBL
treatment for self--energies. Another one is based on complete
$2$nd--order QDT. These approximate schemes significantly reduce the
computational cost, and improve the efficiency for solving the
reduced electronic dynamics. Additional adiabatic approximation is
introduced for the WBL scheme, and the resulting AWBL approximation
turns out the most efficient scheme. Numerical simulations based on
the AWBL scheme demonstrate that first--principles simulation can
indeed be carried out for real devices, and the interesting results
have been obtained.

\begin{acknowledgements}
Support from the Hong Kong Research Grant Council (HKU7009/09P,
7008/08P, 7013/07P, and 604709) is gratefully acknowledged.
\end{acknowledgements}

\appendix

\section{Detailed derivation of HEOM for an open noninteracting system}
\label{ab_deriv_HEOM}
Here we give a direct derivation of HEOM for a noninteracting
system, without resorting to the path--integral formalism for
Fermion operators.
Similarly with Ref.~\onlinecite{Jin08234703}, we introduce the
reservoir $\hat{H}_{\rm B}$--interaction picture, where
$\hat{H}_{\rm B} = \sum_{\alpha}\hat{H}_\alpha$.
In this interaction picture, the total Hamiltonian is
\be
 \tilde{H}_{\rm T}(t) = H_{\rm D} +
 \sum_{\alpha}\tilde{H}_{\alpha {\rm D}}(t).
\ee
Here, $\tilde{H}_{\alpha {\rm D}}(t)=\sum_\mu \big[{\tilde
f}^\dagger _{\alpha \mu}(t)a_\mu + a^\dagger_\mu{\tilde f}_{\alpha
\mu}(t)\big]$ where ${\tilde f}^\dagger _{\alpha \mu }\left(t
\right) =\sum_k \bm t_{\alpha k\mu }\, \tilde d^\dagger _{\alpha
k}(t)$.
The quantum Liouville equation reads
\begin{align}  \label{QLE}
\dot{{\tilde \rho}}_{\rm T}(t) &= -i\big[\tilde{H}_{\rm T}(t),
\tilde{\rho}_{\rm T}(t) \big] \nl &= -i{\cal L}_{\rm D}\,
\tilde{\rho}_{\rm T}(t) - i\sum_{\alpha }{\cal L}_{\alpha {\rm D}}\,
\tilde{\rho}_{\rm T}(t),
\end{align}
where the superoperators are defined as
\begin{align}
 {\cal L}_{\rm D} ~\cdot &\equiv \left[H_{\rm D},\,\cdot \right], \nl
 {\cal L}_{\alpha {\rm D}} ~\cdot &\equiv \big[\tilde{H}_{\alpha {\rm D}}(t),\,\cdot
 \big].
\end{align}
First, we establish the EOM for the RSDM of primary interest, $\bm
\sigma _{\rm D \mu \nu }(t) \equiv \trt [a^\dagger _\nu a_\mu \rho_{\rm
T}(t)] = \trt[a^\dagger_\nu a_\mu \tilde{\rho}_{\rm T}(t)]$.
Following \Eq{QLE}, we have
%
\begin{align}
&\quad i\dot {\bm \sigma}_{\rm D \mu \nu}(t) \nl
 &= \trt [a^\dagger _\nu a_\mu\,{\cal L}_{\rm D}\,\tilde{\rho}_{\rm T}(t)]
+\sum_{\alpha} \trt [a^\dagger _\nu a_\mu\,{\cal L}_{\alpha {\rm
D}}\,\tilde{\rho}_{\rm T}(t)] \nl
&= \trt \left\{ \left[a^\dagger_\nu a_\mu, H_{\rm
D}\right]\tilde{\rho}_{\rm T}(t) \right\} \nl
&\quad +\sum_{\alpha} \trt \big\{ \big[a^\dagger _\nu a_\mu,
\tilde{H}_{\alpha {\rm D}}(t)\big]\tilde{\rho}_{\rm T}(t)\big\} \nl
&= \sum_{\mu'\in {\rm D}} \trt\big[ \big(\bm h_{\rm D \mu \mu'} a^\dagger
_\nu a_{\mu'}- \bm h_{\rm D \mu' \nu} a^\dagger _{\mu'} a_\nu
\big)\tilde{\rho }_{\rm T}(t) \big] \nl
&\quad +\sum_{\alpha}\sum_{\mu'\in {\rm D}} \trt \Big\{\big[-{\tilde
f}^\dagger _{\alpha \mu'}(t)a_\mu\, {\delta}_{\nu \mu'} \nl
&\qquad \qquad \qquad + a^\dagger_\mu{\tilde f}_{\alpha \mu'}(t)
\,\delta_{\mu \mu'}\big]\,\tilde{\rho}_{\rm T}(t)\Big\} \nl
&= \left[\bm h\dd,\bm\sigma\dd(t) \right]_{\mu \nu }-\sum_{\alpha}
\big[\bm \varphi_{\alpha, \mu \nu}(t) - \bm
\varphi^{\dagger}_{\alpha, \mu \nu}(t) \big].
\end{align}
%
Here, the second equality results from trace cycling invariance
property; and the fourth follows the definition of auxiliary RSDM,
$\bm \varphi_{\alpha, \mu\nu}(t) \equiv \trt [{\tilde f}^\dagger
_{\alpha \nu}(t)\, a_\mu{\tilde{\rho}}_{\rm T}(t)]$, and its
Hermitian conjugate $\bm \varphi^\dagger _{\alpha, \mu \nu}(t) =
\trt [a^\dagger_\nu{\tilde f}_{\alpha \mu}(t){\tilde{\rho }}_{\rm
T}(t)]$.

Considering the multiple--frequency--dispersed scheme in
Ref.~\onlinecite{Jin08234703}, we can introduce the frequency
dependence of $\tilde f$ and $\tilde f^\dagger$ from the beginning.
In the reservoir $\hat{H}_{\rm B}$--interaction picture,
\begin{align}
  \dot{\tilde d}^\dagger_{\alpha k}(t) &=
  -i\big[\tilde d^\dagger _{\alpha k}(t), \hat{H}_{\rm B} \big] \nl
 &=i\left[\ep _{\alpha k}+\Delta _\alpha(t) \right]\tilde d^\dagger_{\alpha
 k}(t).
\end{align}
Simple integration leads to
\begin{align}
   \tilde d^\dagger _{\alpha k}(t) &= e^{i\int_{t_0}^t \left[\ep _{\alpha k}
   +\Delta _\alpha(\tau) \right] d\tau}\, \tilde d^\dagger _{\alpha k}, \nl
 {\tilde f}^\dagger_{\alpha \mu}(t) &=\sum_{k\in\alpha} \bm t_{\alpha k\mu}\,\tilde d^\dagger _{\alpha k}\,
 e^{i\int_{t_0}^{t}d\tau \left[\ep _{\alpha k}+\Delta _\alpha(\tau)\right]} \nl
 &= \sum_{k\in\alpha} \int d\omega\, \delta (\omega - \ep _{\alpha k})\,e^{i\omega \left(t-t_0
 \right)}\,
 \bm t_{\alpha k\mu}\, \tilde d^\dagger _{\alpha k} \nl
&\quad \times e^{i\int_{t_0}^{t}d\tau \Delta _\alpha(\tau)}.
\end{align}
Let
\begin{align}
  {\tilde f}^\dagger _{\alpha \mu}(\omega, t)
&=\sum_{k\in\alpha} \delta \left(\omega - \ep _{\alpha k}\right)
e^{i\omega \left(t-t_0 \right)}\, \bm t_{\alpha k\mu}\, \tilde
d^\dagger _{\alpha k} \nl
&\quad \times e^{i\int_{t_0}^{t}d\tau \Delta_\alpha \left(\tau
\right)},
\end{align}
we have ${\tilde f}^\dagger_{\alpha \mu}(t) = \int d\omega {\tilde
f}^\dagger _{\alpha \mu}(\omega, t)$, ${\tilde f}_{\alpha \mu}(t) =
\int d\omega {\tilde f}_{\alpha \mu}(\omega, t)$, and $
\tilde{H}_{\alpha {\rm D}}(t) = \sum_\mu \int d\omega \big[{\tilde
f}^\dagger _{\alpha \mu} (\omega, t)a_\mu + a^\dagger_\mu{\tilde
f}_{\alpha \mu}(\omega, t)\big]$.

Define
\begin{align}
    \bm \varphi _{\alpha, \mu \nu}(\omega, t) &\equiv \trt \big\{\tilde f^\dagger _{\alpha \nu}
(\omega, t)\, a_\mu\, {\tilde{\rho}}_{\rm T}(t)\big\}, \\
    \bm \varphi ^\dagger _{\alpha, \mu \nu}(\omega, t)
&\equiv \trt \big\{a^\dagger_\nu\, {\tilde f}_{\alpha \mu}(\omega,
t)\, {\tilde{\rho}}_{\rm T}(t) \big\}.
\end{align}
They satisfy $\bm \varphi_{\alpha, \mu \nu}(t) = \int d\omega\, \bm
\varphi_{\alpha, \mu \nu}(\omega, t)$ and $\bm
\varphi^\dagger_{\alpha, \mu \nu}(t) = \int d\omega\, \bm\varphi
^\dagger_{\alpha, \mu \nu}(\omega, t)$.
Comparing with the NEGF formalism, it is readily verified that
\be \bm \varphi_{\alpha, \mu \nu} (\omega, t) = -i\sum_{k\in\alpha}
\delta \left(\omega - \ep _{\alpha k}\right) \bm G^<_{\mu \alpha
k}\left(t,t \right) \bm t_{\alpha k\nu}. \label{phi-mu-nu} \ee
The following nonequilibrium Green's function is defined on the
Keldysh contour, where the time variable goes from $- \infty$ to $+
\infty$ then back to $- \infty$:
\begin{align}
\bm G_{\mu \alpha k}\left(\tau, \tau'\right) &\equiv -i\, \trt
\left\{T_C \big[a_\mu \left(\tau  \right)d^\dagger _{\alpha
k}\left(\tau' \right) \big] \rho_{\rm T}\left(t_0 \right)\right\}
\nl
&= \sum_{\nu\in{\rm D}} \int_{C} d\tau_1 \bm G_{\mu \nu}\left(\tau,
\tau_1\right) \bm t^*_{\alpha k\nu}\,g_{\alpha k}\left(\tau_1,
\tau'\right), \label{NEGF1}
\end{align}
where the second equality follows the derivation for Eq.~(B7) in
Appendix~B of Ref.~\onlinecite{Jauho}.
With the Langreth's analytic continuation rules,\cite{Langreth} we
have
\begin{align}
\bm G^<_{\mu \alpha k} \left(t, t\right) &= \sum_{\nu\in {\rm D}}
\bm t^*_{\alpha k \nu} \int_{-\infty}^{\infty} dt_1 \left[ \bm
G^r_{\mu \nu} \left(t, t_1\right) g^<_{\alpha k} \left(t_1, t\right)
\right. \nl &\quad \left. +\, \bm G^<_{\mu \nu} \left(t, t_1\right)
g^a_{\alpha k} \left(t_1, t\right) \right]. \label{g-d-less}
\end{align}
Inserting \Eq{g-d-less} into \Eq{phi-mu-nu} leads to
%
\begin{widetext}
\begin{align}
\bm \varphi _{\alpha, \mu \nu} (\omega, t) &=
-i\sum_{k\in\alpha}\sum_{\nu'\in {\rm D}} \bm t^*_{\alpha k
\nu'}\,\bm t_{\alpha k \nu}\, \delta(\omega - \ep_{\alpha k})
\int_{-\infty}^{\infty}dt_1 \left[\bm G^r_{\mu \nu'}(t, t_1)\,
g^<_{\alpha k}(t_1, t) + \bm G^<_{\mu \nu'}(t, t_1)\, g^a_{\alpha
k}(t_1, t)\right] \nl
&=-i\sum_{\nu'\in {\rm D}}\int_{-\infty}^{\infty} dt_1 \left[\bm
G^r_{\mu \nu'} \left(t,t_1\right) \bm \Sigma^<_{\alpha,
\nu'\nu}\left(t_1,t;\omega \right) + \bm G^<_{\mu \nu'} \left(t,
t_1\right) \bm \Sigma^a_{\alpha, \nu' \nu} \left(t_1, t;
\omega\right) \right] \nl
&=i\sum_{\nu'\in{\rm D}}\int_{-\infty}^{t} dt_1 \left[\bm G^<_{\mu
\nu'} \left(t, t_1\right) \bm \Sigma^>_{\alpha, \nu' \nu} \left(t_1,
t; \omega\right) - \bm G^>_{\mu \nu'} \left(t, t_1 \right) \bm
\Sigma^<_{\alpha, \nu' \nu} \left(t_1, t; \omega\right) \right],
\label{phi-alpha-mu-nu}
\end{align}
\end{widetext}
where the third equality employs the relation
\be \bm X^{r,a}\left(t,\tau\right)=\pm \vartheta
\left[\pm\left(t-\tau\right)\right]\left[\bm X^>(t,\tau)- \bm
X^<(t,\tau)\right] \ee
with $\bm X$ being $\bm G\dd$ or $\bm \Sigma_\alpha$.
The frequency dispersed self--energies are defined by
\be \bm\Sigma^x_{\alpha, \nu' \nu}(t_1, t; \omega) \equiv
\sum_{k\in\alpha} \bm t^*_{\alpha k \nu'} \bm t_{\alpha k\nu} \,
\delta(\omega - \ep_{\alpha k})\, g^x_{\alpha k}(t_1, t),
\label{selfeng} \ee
where $x = r$, $a$, $<$, and $>$. They satisfy
\begin{align}
\bm \Sigma^<_{\alpha, \nu' \nu}(t_1, t) &= \int d\omega \, \bm
\Sigma^<_{\alpha, \nu' \nu}(t_1, t; \omega) \nl
&= i\sum_{k\in\alpha} \bm t^*_{\alpha k \nu'} \bm t_{\alpha k\nu}
\int d\omega\, \delta(\omega - \ep _{\alpha k}) f_\alpha \left(\ep
_{\alpha k}\right) \nl
&\quad \times e^{i\omega \left(t-t_1 \right)}
e^{i\int_{t_1}^{t}d\tau \Delta_\alpha \left(\tau \right)} \nl
&=i\int d\omega\, \bm\Lambda_{\alpha, \nu' \nu}(\omega)
f_\alpha(\omega)\, e^{i\omega \left(t-t_1 \right)} \nl
&\quad \times e^{i\int_{t_1}^{t}d\tau \Delta _\alpha \left(\tau
\right)}.
\end{align}

Next, we establish the EOM for the first--tier auxiliary RSDM,
$\bm\varphi_{\alpha, \mu \nu}(\omega, t)$, as follows,
\begin{align}
&\qquad {\dot{\bm \varphi}}_{\alpha, \mu \nu}(\omega, t)\nl
&= \trt \big \{\dot{\tilde f}^\dagger _{\alpha \nu }(\omega,
t)\,a_\mu\, {\tilde{\rho }}_{\rm T}(t) + \tilde f^\dagger _{\alpha
\nu} (\w, t)a_\mu\,{\dot{\tilde{\rho}}}_{\rm T}(t)\big\} \nl
&=i\left[\omega+\Delta_\alpha(t) \right]\bm \varphi _{\alpha, \mu
\nu}(\omega, t) \nl &\quad +\trt \big \{\tilde f^\dagger _{\alpha
\nu }(\omega, t)\, a_\mu \big[-i{\cal L}\dd\,{\tilde{\rho }}_{\rm
T}(t)\big] \big\} \nl
&\quad +\sum_{\alpha'}\trt \big\{\tilde f^\dagger _{\alpha \nu }(\w,
t)\,a_\mu \big[-i{\cal L}_{\alpha' \rm D}\,{\tilde{\rho }}_{\rm
T}(t)\big] \big\} \nl
&= i\left[\omega+\Delta _\alpha (t) \right]\bm \varphi _{\alpha, \mu
\nu} (\omega, t) \nl
&\quad - i\,\trt \big\{\big[\tilde f^\dagger _{\alpha \nu } (\w,
t)\,a_\mu, H\dd\big]\tilde{\rho }_{\rm T} (t)\big\} \nl
&\quad - i\sum_{\alpha'}\trt \big\{ \big[\tilde f^\dagger _{\alpha
\nu } (\omega, t)\,a_\mu, \tilde{H}_{\alpha' {\rm D}}(t)
\big]\tilde{\rho }_{\rm T}(t)\big\} \nl
&= i\left[\omega+\Delta _\alpha (t) \right] \bm \varphi _{\alpha,
\mu \nu }(\omega, t)\nl
&\quad - i\sum_{\mu'\in{\rm D}} \bm h_{\rm D \mu \mu'}\,\trt \big\{\tilde
f^\dagger _{\alpha \nu } (\w, t)\,a_{\mu'}\,\tilde{\rho }_{\rm T}(t)
\big\} \nl
&\quad - i \sum_{\alpha'} \int d\omega' \trt \big\{\tilde f^\dagger
_{\alpha \nu } (\omega, t)\,\tilde f_{\alpha'\mu }(\w',
t)\,\tilde{\rho }_{\rm T}(t) \big\} \nl
&\quad + i\sum_{k\in\alpha}\sum_{\mu'\in{\rm D}}\trt \big\{\bm
t_{\alpha k\nu}\, \bm t^*_{\alpha k \mu'} \delta(\omega - \ep
_{\alpha k})\, a^\dagger_{\mu'}\, a_\mu \, \tilde{\rho }_{\rm
T}(t)\big\} \nl
&=i\left[\omega+\Delta_\alpha(t) \right] \bm \varphi _{\alpha, \mu
\nu }(\omega, t) \nl
&\quad - i\sum_{\mu'\in {\rm D}} \big[\bm h_{\rm D \mu \mu'} \,\bm \varphi
_{\alpha, \mu'\nu}(\omega, t) - \bm\sigma_{\mu\mu'}(t)\,
\bm\Lambda_{\alpha, \mu'\nu}(\w) \big] \nl
&\quad -i \sum_{\alpha'} \int d\omega'\, \trt \big \{\tilde
f^\dagger _{\alpha \nu}(\omega, t)\, \tilde f_{\alpha'\mu}(\omega',
t)\, \tilde{\rho }_{\rm T}(t) \big \}. \label{eom-1st-rsdm}
\end{align}

Comparing with the NEGF formalism, it is readily verified that
\begin{align} \label{2aux}
&\qquad \trt \big\{\tilde f^\dagger _{\alpha \nu }(\omega,
t)\,\tilde f_{\alpha'\mu }(\w', t)\, \tilde{\rho }_{\rm T}(t)\big \}
\nl &=\sum_{k\in\alpha}\sum_{k'\in\alpha'} \delta(\omega -
\ep_{\alpha k})\, \delta(\omega' - \ep _{\alpha' k'})\, \bm
t_{\alpha k\nu} \bm t^*_{\alpha' k' \mu} \nl
&\qquad \times \trt \big\{\tilde d^\dagger _{\alpha k}\,\tilde
d_{\alpha'k'}\, \tilde{\rho }_{\rm T}(t) \big \} \nl
&=-i \sum_{k\in\alpha}\sum_{k'\in\alpha'} \delta(\omega - \ep
_{\alpha k})\,\delta(\omega' - \ep _{\alpha' k'})\, \bm t_{\alpha
k\nu} \bm t^*_{\alpha' k' \mu} \nl
&\qquad \times \bm G^<_{\alpha'k', \alpha k}(t, t).
\end{align}
Generally, the NEGF on the same Keldysh contour satisfies
\begin{align}  \label{NEGF2}
\bm G_{\alpha'k', \alpha k}\left(\tau ,\tau'\right) &=-i\,\trt
\big\{T_C \big[d_{\alpha'k'}(\tau)\,d^\dagger_{\alpha k}(\tau')
\big] \rho_{\rm T}(t_0) \big\} \nl
&= \delta_{\alpha \alpha'} \delta_{kk'}\, g_{\alpha k}(\tau, \tau')
\nl
&\quad + \sum_{\mu_1 \mu_2\in{\rm D}} \int_C d\tau_1 \int_C d\tau_2
\, g_{\alpha' k'}\left(\tau ,\tau_1\right) \bm t_{\alpha'k' \mu_1}
\nl &\quad \times \bm G_{\mu_1 \mu_2}\left(\tau_1, \tau_2\right) \,
\bm t^*_{\alpha k \mu_2} \,g_{\alpha k}\left(\tau_2,\tau'\right),
\end{align}
where the same technique in Ref.~\onlinecite{Jauho} adopted for
derivation of \Eq{NEGF1} is used.
At $\tau=\tau'=t$ the lesser component of Green's function for the
uncoupled lead is $g^<_{\alpha k}\left(t,t\right)=g^<_{\alpha
k}(t-t) = i f_{\alpha} \left(\ep _{\alpha k}\right)$. Inserting the
first term on RHS of \Eq{NEGF2} into \Eq{2aux} leads to
\begin{align}
&\quad -i\,\sum_{\alpha'}\sum_{k\in\alpha}\sum_{k'\in\alpha'} \int
d\omega'\, \delta(\omega - \ep _{\alpha k})\, \delta(\omega' - \ep
_{\alpha' k'}) \nl
&\qquad \qquad \qquad \qquad \times \bm t_{\alpha k\nu}\,\bm
t^*_{\alpha' k' \mu}\delta_{\alpha \alpha'} \delta_{kk'}\, \big[i
f_\alpha\left(\ep _{\alpha k}\right) \big] \nl
&= \bm \Lambda_{\alpha, \mu \nu}(\omega) \, f_\alpha(\omega).
\end{align}
Inserting the second term on RHS of \Eq{NEGF2} into \Eq{2aux} while
using \Eq{selfeng}, we arrive at a term on the RHS 
of \Eq{eom-1st-rsdm}
%
\begin{align}
\bm \varphi_{\alpha\alpha', \nu\mu}(\w, \w', t) &= -i
\!\!\sum_{\mu_1 \mu_2\in{\rm D}} \int_C d\tau_1 \int_C d\tau_2 \big[
\bm \Sigma_{\alpha', \mu \mu_1}(t, \tau_1; \omega') \nl & \times \bm
G_{\mu_1 \mu_2 }(\tau_1 , \tau_2) \bm \Sigma_{\alpha, \mu_2
\nu}(\tau_2, t; \omega) \big]^<. \label{1def}
\end{align}
%
It turns out that the EOM for the first--tier auxiliary RSDM,
$\bm\varphi _{\alpha, \mu \nu}(\omega, t)$, is formally identical
to that given by path--integral formulation in
Ref.~\onlinecite{Jin08234703}. Combining \Eqs{2aux} through
(\ref{1def}), \Eq{eom-1st-rsdm} finally reads:
%
\begin{align}
i{\dot{\bm \varphi}}_{\alpha, \mu \nu}(\omega, t)
&=-\left[\omega+\Delta _\alpha(t) \right] \bm \varphi _{\alpha, \mu
\nu }(\omega, t) \nl
&\quad + \sum_{\mu'\in{\rm D}} \bm h_{\rm D \mu \mu'} \bm \varphi
_{\alpha, \mu' \nu }(\omega, t) \nl
&\quad -\sum_{\mu'\in{\rm D}} \bm \sigma_{\rm D \mu \mu'}(t) \bm
\Lambda_{\alpha,\mu'\nu}(\w) + f_\alpha(\w)\bm \Lambda_{\alpha, \mu
\nu }(\w) \nl
&\quad +\sum_{\alpha'} \int d\omega'\, \bm
\varphi_{\alpha\alpha',\nu \mu }(\w, \w', t).
\end{align}
%
Here the compact definition  (\ref{1def}) establishes the natural
connection between the second--tier auxiliary
RSDM and NEGF quantities.
Again, using the Langreth's analytic continuation
rules,\cite{Langreth} we can immediately write down a real time
expression for \Eq{1def} as follows, which looks however more
complicated.
\begin{widetext}
\begin{align}
&\quad \bm \varphi_{\alpha\alpha',\nu\mu }(\w, \w', t) \nl
&= i\sum_{\mu_1 \mu_2\in{\rm D}} \bigg\{ \int_{-\infty}^{t} dt_2\,
\bigg[\int_C d\tau_1 \, \bm\Sigma_{\alpha', \mu
\mu_1}(t,\tau_1;\omega')\, \bm G_{\mu_1 \mu_2 }(\tau_1, t_2)
\bigg]^< \,\bm \Sigma^>_{\alpha, \mu_2 \nu}(t_2, t; \omega) \nl
&\qquad \quad -\int_{-\infty}^{t} dt_2\, \bigg[\int_C d\tau_1 \bm
\Sigma_{\alpha', \mu \mu_1}(t,\tau_1; \omega') \bm G_{\mu_1
\mu_2}(\tau_1, t_2)\bigg]^> \, \bm\Sigma^<_{\alpha, \mu_2 \nu}(t_2,
t; \omega) \bigg\} \nl
&=i\sum_{\mu_1 \mu_2\in{\rm D}} \bigg\{ \int_{-\infty}^{t} dt_2
\int_{-\infty}^{t} dt_1 \left[\bm\Sigma^<_{\alpha', \mu \mu_1}
(t,t_1;\omega') \bm G^a_{\mu_1 \mu_2 }(t_1, t_2) + \bm
\Sigma^r_{\alpha', \mu \mu_1}(t, t_1; \omega') \bm G^<_{\mu_1
\mu_2}(t_1, t_2) \right] \,\bm\Sigma^>_{\alpha, \mu_2 \nu}(t_2, t;
\omega) \nl
&\qquad \quad- \int_{-\infty}^{t} dt_2 \int_{-\infty}^{t} dt_1
\left[\bm \Sigma^r_{\alpha', \mu \mu_1}(t, t_1; \omega') \bm
G^>_{\mu_1 \mu_2}(t_1, t_2) + \bm\Sigma^>_{\alpha', \mu \mu_1} (t,
t_1; \omega') \bm G^a_{\mu_1 \mu_2}(t_1, t_2)\right] \,\bm
\Sigma^<_{\alpha, \mu_2 \nu}(t_2, t; \omega) \bigg\}.
\end{align}
\end{widetext}
The last equality recovers Eqs.~(22) and (24) in
Ref.~\onlinecite{croy2}.
At $t\rightarrow -\infty$ the device and leads are fully decoupled,
\ie $\tilde \rho_{\rm T}(-\infty) = \rho_{\rm B} \otimes
\tilde{\rho}\dd(-\infty)$.
Noting that
\begin{align}
&\quad \bm\Lambda_{\alpha, \mu \nu}(\w)f_\alpha(\w) \nl
&=\int d\omega' \,\trt \big[\tilde f^\dagger _{\alpha \nu } (\omega,
t)\tilde f_{\alpha'\mu} (\w', t) \, \tilde \rho_{\rm T}(-\infty)
\big],
\end{align}
we can have a most simple and compact
expression of the second--tier auxiliary RSDM as follows,
\begin{align}
&\quad \bm \varphi_{\alpha\alpha',\nu\mu}(\omega, \omega',t) \nl
&= \trt \big\{\tilde f^\dagger _{\alpha \nu}(\omega, t)\, \tilde
f_{\alpha'\mu } (\w', t) \left[\tilde{\rho }_{\rm T}(t) -\tilde
\rho_{\rm T}(-\infty)\right] \big\}.  \label{2def}
\end{align}
Based on \Eq{2def}, it is easier to obtain the EOM for the
second--tier auxiliary RSDM, $\bm\varphi_{\alpha\alpha', \nu\mu
}(\w, \w', t)$, as follows.
\begin{widetext}
\begin{align}
&\quad \dot{\bm\varphi}_{\alpha\alpha', \nu\mu}(\omega, \omega', t)
\nl
&=i\left[\omega+\Delta_\alpha(t) \right] \bm\varphi_{\alpha
\alpha',\nu\mu}(\w, \w', t) -i\left[\omega'+\Delta _{\alpha'} (t)
\right] \bm\varphi_{\alpha\alpha', \nu\mu}(\w, \w', t) +\trt
\big[\tilde f^\dagger _{\alpha \nu }(\omega, t)\,\tilde
f_{\alpha'\mu }(\w', t)\,\dot{\tilde{\rho }}_{\rm T}(t) \big] \nl
&=i\left[\omega+\Delta_\alpha(t)-\omega' -\Delta _{\alpha'}(t)
\right] \bm\varphi_{\alpha\alpha', \nu\mu}(\w, \w', t) + \trt
\big\{\tilde f^\dagger _{\alpha \nu }(\omega, t)\,\tilde
f_{\alpha'\mu }(\w', t)\,\left[-i{\cal L}\dd\,\tilde{\rho}_{\rm
T}(t)\right] \big\} \nl
&\quad +\sum_{\alpha_1}\trt \big\{\tilde f^\dagger_{\alpha \nu}
(\omega, t)\tilde f_{\alpha'\mu }\left(\omega', t
\right)\,\left[-i{\cal L}_{\alpha_1 {\rm D}}\,\tilde{\rho }_{\rm
T}(t) \right] \big\} \nl
&=i\left[\omega+\Delta_\alpha(t) -\omega'-\Delta _{\alpha'}(t)
\right] \bm \varphi_{\alpha\alpha',\nu\mu}(\w, \w', t) -i\,\trt \big
\{ \big[\tilde f^\dagger _{\alpha \nu }(\omega, t)\,\tilde
f_{\alpha'\mu }(\w', t), H\dd \big]\tilde{\rho}_{\rm T}(t) \big\}
\nl
&\quad -i\sum_{\alpha_1} \trt \big\{ \big[\tilde f^\dagger _{\alpha
\nu }(\omega, t)\, \tilde f_{\alpha'\mu }\left(\omega', t\right),
\tilde{H}_{\alpha' {\rm D}}(t)\big]\,\tilde{\rho }_{\rm T}(t) \big\}
\nl
&=i\left[\omega+\Delta _\alpha (t)-\omega'-\Delta _{\alpha'} (t)
\right] \bm \varphi_{\alpha\alpha', \nu\mu}(\omega, \omega',t) \nl
&\quad -i\sum_{\alpha_1}\sum_{\mu'\in{\rm D}}\sum_{k_1\in\alpha_1}
\sum_{k'\in\alpha'} \trt \big[\tilde f^\dagger _{\alpha
\nu}\left(\omega, t \right) a_{\mu'} \delta_{\alpha' \alpha_1}
\delta_{k'k_1} \delta(\omega'-\ep _{\alpha' k'})\, \bm t_{\alpha_1
k_1 \mu'}\, \bm t^*_{\alpha' k' \mu}\, \tilde{\rho}_{\rm T}(t) \big]
\nl
&\quad + i\sum_{\alpha_1}\sum_{\mu'\in{\rm D}}
\sum_{k_1\in\alpha_1}\sum_{k\in\alpha} \trt \big[a^\dagger _{\mu'}
\tilde f _{\alpha \mu}(\w', t)\, \delta_{\alpha \alpha_1}
\delta_{kk_1} \delta(\omega-\ep _{\alpha k})\, \bm t_{\alpha_1 k_1
\nu}\, \bm t^*_{\alpha k \mu'}\, \tilde{\rho}_{\rm T}(t) \big] \nl
&= i \left[\omega+\Delta_\alpha(t) -\omega'-\Delta _{\alpha'}(t)
\right] \bm \varphi_{\alpha\alpha', \nu\mu}(\w, \w', t)
-i\sum_{\mu'\in{\rm D}} \bm \Lambda_{\alpha',\mu \mu'} (\omega') \bm
\varphi_{\alpha, \mu' \nu}(\omega, t) +i\sum_{\mu'\in{\rm D}} \bm
\varphi^\dagger_{\alpha',\mu \mu'}(\omega', t) \bm \Lambda_{\alpha,
\mu' \nu}(\w).
\end{align}
\end{widetext}
The last equality recovers Eq.~(6.6) in
Ref.~\onlinecite{Jin08234703}.

\section{Remarks on Landauer--B\"{u}ttiker formula for steady current} \label{sstate}


Under an external voltage applied to electrode $\alpha$,
a homogeneous energy shift $\Delta_\alpha$ is resulted
for all energy levels, and the self--energy is thus
 \be
   \bm\Sigma^x_\alpha(t,\tau) = \exp\left[ -i\int_\tau^t
   \Delta_\alpha(\zeta) d\zeta \right]
   \tilde{\bm\Sigma}^x_\alpha(t-\tau), \label{self-x}
 \ee
where $x=r,a,<$, and $>$; and the tilde symbol denotes quantities of
ground/equilibrium--state in absence of any voltage, where
translational invariance in time applies.

For the composite system to approach a steady state as $t
\rightarrow +\infty$, it is necessary to have $\Delta_\alpha^\infty
\equiv \Delta_\alpha(t\rightarrow +\infty)$ approaching a constant
for any $\alpha$. Based on Riemann--Lebesgue lemma, $\bm
G\dd(t,\tau) = \bm \Sigma_\alpha(t,\tau) = 0$ as $t \rightarrow
+\infty$ and $\tau$ remains finite. For $t,\tau \rightarrow
+\infty$, $\bm G\dd(t,\tau) = \bm G\dd(t-\tau)$ and $\bm
\Sigma_\alpha(t,\tau) = \bm \Sigma_\alpha(t-\tau)$, \emph{i.e.},
translation invariance in time is retrieved at steady state.

The NEGF formalism gives steady current through electrode $\alpha$,
$J_\alpha^\infty \equiv J_\alpha(t\rightarrow +\infty)$, as follows,
 \begin{align}
   J_\alpha^\infty &= -
   {\rm tr}\dd \left[ \bm Q_\alpha(t\rightarrow +\infty) \right]\nl
   &= \sum_{\alpha' \neq \alpha} \int d\eps \left[f_\alpha(\eps) -
   f_{\alpha'}(\eps)\right] T_{\alpha\alpha'}(\eps), \label{landauer} \\
   T_{\alpha\alpha'}(\eps) &= 2\pi\,
   {\rm tr}\dd \Big[ \bm G\dd^r(\eps)
   \bm \Lambda_{\alpha'}(\eps^\infty_{\alpha'})\bm G\dd^a(\eps)
   \bm \Lambda_{\alpha}(\eps^\infty_{\alpha}) \Big]. \label{t-of-e}
 \end{align}
Here, $\eps^\infty_\alpha \equiv \eps - \Delta^\infty_\alpha$, and
$T_{\alpha\alpha'}(\eps)$ is the KS transmission coefficient between
electrodes $\alpha$ and $\alpha'$. Equations~(\ref{landauer}) and
(\ref{t-of-e}) correspond formally to the Landauer--B\"{u}ttiker
formula. Equation~(\ref{landauer}) has also been reached in the
framework of HEOM--QDT, see Ref.~\onlinecite{Jin08234703}. It is
worth emphasizing that the exact XC potential of TDDFT is in
principle frequency--dependent.

Equations~(\ref{landauer}) and (\ref{t-of-e}) have been widely
employed in the DFT--NEGF approach to simulate steady current
through molecular devices and structures.
%
%
The exact XC potential may depend explicitly on the time evolution
history of electron density in TDDFT, while it is determined
exclusively by $\rho(\bm r, t\rightarrow+\infty)$ in stationary DFT.
Under the same time--dependent applied voltage, DFT and TDDFT
frameworks result in explicitly different steady current \emph{if}
and \emph{only if} a nonadiabatic (or frequency--dependent) XC
potential is adopted in TDDFT framework. In other words, DFT--NEGF
approach is regarded as an adiabatic version of TDDFT--NEGF.

In principle, the steady current may also vary upon different
time--dependent applied voltages, even if the voltage amplitudes are
same at $t \to +\infty$. For instance, there may be multiple steady
states corresponding to the multiple stationary solutions of the
TDDFT--EOM, \Eq{eom0}. The actual steady state reached at
$t\to+\infty$ is thus completely determined by the initial electron
density, and the history of externally applied voltage.

It has been shown that bound states in the device region would give
rise to persistent oscillating current across the device--electrode
interfaces.\cite{Ste07195115, Kho094535} In such cases, the
Landauer--B\"{u}ttiker formula becomes inadequate for describing the
long--time limit of transient current. In fact, these bound states
are physically decoupled with the rest of composite system, and form
an isolated subsystem. When this isolated subsystem possesses
nontrivial component (such as tail of wavefunction) inside the $\bm
r$--space of electrodes, the persistent oscillating current would
arise, as an electron transfers across the device--electrode
interfaces, but remains within the subsystem. It is thus inferred
that the magnitude of oscillating current would diminish as the
device region is enlarged by pushing the interfaces (where the
currents are measured) deeper into the electrodes. The oscillating
current would vanish completely with all the bound states fully
accommodated in the device region.

%
%

%

\end{document}